\documentclass[12pt]{article}

\usepackage{amssymb,amsmath,mathbbol,mathrsfs}
\usepackage[usenames,dvipsnames]{color}
\usepackage[version=3]{mhchem}
\usepackage{hyperref}
\usepackage[table]{xcolor}
\usepackage{multirow,bigdelim}
\usepackage{stmaryrd}
\usepackage{authblk}
\usepackage{framed}
\usepackage{empheq}
\usepackage{slashed}
\usepackage{hyphenat}
\usepackage{cite}
\usepackage{stackrel}

\usepackage{epsfig}

\usepackage{tikz}
\usepackage{tikz-3dplot}
\usetikzlibrary{calc}
\usetikzlibrary{shapes.geometric,patterns} 
\usetikzlibrary{tikzmark}

\usepackage{marginnote}    

\usepackage[left=0.5in,right=1in,top=1in,bottom=1in,marginparwidth=1.5in, marginparwidth=2.5cm, marginparsep=.2cm]{geometry}       

\hypersetup{
pdfstartview = {FitH},
}

\def\UQ{{\cU_{q}(\su(2))}}

\def\poi#1{\{ #1 \}}
\newcommand{\be}{\begin{equation}}
\newcommand{\ee}{\end{equation}}
\newcommand{\beq}{\begin{equation}}
\newcommand{\eeq}{\end{equation}}
\newcommand{\bes}{\begin{eqnarray}}
\newcommand{\ees}{\end{eqnarray}}
\newcommand{\bqa}{\begin{eqnarray}}
\newcommand{\eqa}{\end{eqnarray}}
\newcommand{\bea}{\begin{eqnarray}}
\newcommand{\eea}{\end{eqnarray}}

\newcommand{\R}{\mathbb{R}}
\newcommand{\C}{\mathbb{C}}

\newcommand{\su}{\mathfrak{su}}
\newcommand{\so}{\mathfrak{so}}

\newcommand{\SO}{\mathrm{SO}}

\newcommand{\an}{\mathfrak{an}}

\newcommand{\Sp}{\mathcal{J}}
\newcommand{\Pp}{\mathcal{P}}
\newcommand{\tPp}{\mathcal{P}'}
\newcommand{\tSp}{\mathcal{J}'}
\newcommand{\tP}{{P'}}

\newcommand{\cA}{{\cal A}}
\newcommand{\cD}{{\cal D}}

\newcommand{\cF}{{\cal F}}

\newcommand{\cG}{{\cal G}}

\newcommand{\cJ}{{\cal J}}

\newcommand{\cP}{{\cal P}}

\newcommand{\mF}{{\mathfrak{F}}}
\newcommand{\mT}{{\mathfrak{T}}}

\def\hell{\widehat{\ell}}

\def\demi{{\frac{1}{2}}}
\newcommand{\cL}{{\cal L}}
\newcommand{\act}{{\,\triangleright\,}}

\def\ovo{{\overline{\omega}}}
\def\ove{{\overline{\textbf{e}}}}

\def\ip{\lrcorner\,}

\newcommand{\cU}{{\mathcal U}}

\newtheorem{theorem}{Theorem}

\newtheorem{proposition}{Proposition}

\def\mF{{\mathfrak{F}}}
\def\cs{{\mathfrak{g}}}

\def\mr{{\mathfrak{r}}}

\def\mk{\gamma}

\def\ds{{\mathfrak{d}}}
\def\DS{{\mathfrak{D}}}

\def\sl{{\mathfrak{sl}}}

\newcommand{\la}{\left\langle}
\newcommand{\ra}{\right\rangle}

\def\tell{\tilde{\ell}}

\def\one{{\bf 1}}

\def\pa{\partial}
\def\ka{{\kappa}}

\def\hJ{{\widehat J}}
\def\hP{{\widehat  P}}
\def\tom{ \omega}
\def\om{{ \omega}}

\def\cop{{\Delta}}
\def\ot{{\otimes}}
\def\dr{{\,\rightarrow\,}}
\def\mone{^{-1}}

\def\hell{\widehat{\ell}}

\def\demi{{\frac{1}{2}}}
\def\ovo{{\overline{\omega}}}
\def\ove{{\overline{\textbf{e}}}}

\DeclareMathOperator{\SU}{SU}
\DeclareMathOperator{\AN}{AN}

\DeclareMathOperator{\SL}{SL}

\def\poi#1{\{ #1 \}}
\def\com#1{[ #1 ]}

\def\nn{\nonumber}

\def\bfJ{{{\bf J}} }
\def\bfP{{{\bf P}} }

\def\Dr{{\Delta} }
\def\tl{{\tilde \ell} }
\def\th{{\tilde h} }

\def\Dl{{\underline{\Delta} }}

\newcommand{\rd}{\mathrm{d}}
\newcommand{\rD}{{\mathrm{d}_A}}

\def\Rm{r_-}
\def\Rp{r_+}

\newcommand{\fin}{}

\usepackage{sectsty}
\allsectionsfont{\sffamily\mdseries\upshape} 
\usepackage{tocloft}

\makeatletter
\renewenvironment{abstract}{%
    \if@twocolumn
      \section*{\abstractname}%
    \else 
      \begin{center}%
        {\bfseries\sffamily\abstractname\vspace{\z@}}
      \end{center}%
      \quotation
    \fi}
    {\if@twocolumn\else\endquotation\fi}
\makeatother

\hypersetup{
	colorlinks=true,         
	linkcolor=purple,          
	citecolor=red,        
	urlcolor=Blue            
}

\begin{document}

\title{\sffamily  Origin of the quantum group symmetry in  3d quantum gravity }

\author[1,2]{\sffamily Ma\"it\'e Dupuis\thanks{mdupuis@perimeterinstitute.ca}}
\author[1]{\sffamily Laurent Freidel\thanks{lfreidel@perimeterinstitute.ca}}
\author[2]{\sffamily Florian Girelli\thanks{florian.girelli@uwaterloo.ca}}
\author[2]{\sffamily Abdulmajid Osumanu \thanks{a3osumanu@uwaterloo.ca}}
\author[2]{\sffamily Julian Rennert \thanks{jrennert@uwaterloo.ca}}
\affil[1]{\small Perimeter Institute for Theoretical Physics, 31 Caroline St. N., Waterloo, ON N2L2Y5, Canada}
\affil[2]{\small Department of Applied Mathematics, University of Waterloo, 200 University Avenue West, Waterloo, Ontario, Canada, N2L 3G1}

\maketitle

\abstract{  It is well-known that quantum groups are relevant to describe the quantum regime of 3d gravity. They encode a deformation of the gauge symmetries  parametrized by the value of the cosmological constant. They appear as a form of regularization either through the quantization of the Chern-Simons formulation or the state sum approach of Turaev-Viro. 
Such deformations are perplexing from a continuum and classical picture since the action is defined in terms of undeformed gauge invariance. We present here a novel way to derive from first principles and from the classical action such quantum group deformation. The argument relies on two main steps. First we perform a canonical transformation, which 
deforms the gauge invariance and the boundary symmetries, and makes them  depend on the cosmological constant. Second we implement a discretization procedure relying on a truncation of the degrees of freedom from the continuum.   
}



\tableofcontents

\section*{Introduction}
\label{sec:intro}
When constructing a quantum theory, it is essential to identify the system's relevant symmetries. Symmetries provide, thanks to Noether's theorem \cite{Olver}, a non-perturbative handle which enables us to limit the quantization ambiguities, for example, by demanding that such symmetries are preserved upon quantization. They permit a powerful organization of the spectra by allowing the quantum states to form a representation of the symmetry group. 
 
 For gauge theories such as gravity, it appears that this powerful tool is not available.
 Indeed, it is often believed that there are no symmetries in gravity, only gauge invariances.
 This leaves no means to use the power of having non-trivial conserved charges. Gauge invariances are conventionally understood \cite{Rozali} to be mere redundancies of the parametrization and, therefore, cannot help us organize the quantum spectra. Physical states cannot be distinguished or labeled by the canonical generators associated with gauge invariance since, by definition, they vanish on all physical states.
 We have a state of complete degeneracy,  which is another expression of the celebrated problem of time \cite{Isham:1992ms}, and this is the main reason behind the challenge of constructing a theory of quantum gravity.
 
 Although there is no doubt that gauge invariance implies redundancy of the parameterization, there is a lingering sense that there is more to it \cite{Rovelli:2013fga}. After all, different formulations of gravity,
 such as canonical formulation \cite{Arnowitt:1962hi}, metric formulation \cite{Wald:1984rg}, tetrad formulation \cite{Trautman:2006fp}, teleparallel formulation \cite{tele}, shape dynamics \cite{Barbour:2011dn}, etc., possess different levels of redundancies and seem to present different advantages. Moreover, one seldom studies gravity in a fully gauge fixed form, such as \cite{Grant:2009zz, Verlinde:1991iu},  which would be the most natural and beneficial option if redundancy was all there is to gauge invariance.

It is therefore natural to wonder whether there can be some other types of ``hidden'' symmetries that could be essential in the construction of the quantum theory, and whether such hidden symmetries could entertain a profitable relationship with the notion of gauge invariance? 
Critical examples of hidden symmetries in field theories are dualities \cite{Seiberg}, which are not manifest in the bulk Lagrangian. Other examples of hidden symmetries are dynamical symmetries \cite{Faddeev} that arise in integrable systems. 

One of the first and strongest indications that  there are such  ``hidden'' symmetries in gravity comes from the  Turaev-Viro (TV) model \cite{TV}, which is an expansion of the Ponzano-Regge \cite{Ponzano-Regge} model.  Indeed, in the presence of a cosmological constant, the quantum gravity partition function, can be constructed in terms of 
spin network states satisfying the intertwining properties of quantum groups \cite{Chari:1994pz, Kassel}.
The TV model  provides a discretization of the gravity path integral.
This discretization satisfies two fundamental properties: First, each building block, given by the quantum group 6j symbol, is related in the limit of small Planck constant to the exponential of the classical gravity action \cite{Taylor}. Second, the partition function is invariant under refinement hence defines a continuum theory.
The puzzle comes from the fact that there seems to be no sign of quantum groups in the continuum theory, and quantum groups seem to appear  only after discretization and quantization.

Other mathematical justifications for quantum group symmetries in the context of 3d quantum gravity also originate from the fact that one can relate the TV model 
 to the quantization of Chern-Simons  (CS)
 \cite{ROBERTS, Freidel:2004nb, alex2010, turaev2010}, and then prove that quantum groups 
 appear in the definition of the quantum CS theory.
For instance, the conjecture that quantum groups enter the construction of the CS partition function was first made by Witten \cite{Witten:1988hc} and proven by Reshetikhin-Turaev \cite{RT}. Another important evidence comes from the construction by Fock and Roskly \cite{Fock:1998nu} of a discrete version of the CS phase space, which includes from the get-go arbitrary sets of classical R-matrices.
The quantization of this discrete phase, in terms of quantum groups, was achieved by Alekseev et al.    \cite{Alekseev:1994au, Alekseev:1994pa}.
These approaches are top-down in the sense that quantum groups are postulated in the construction of partition functions or states or algebras and then justified by the consistency of their mathematical properties \emph{but not derived from first principles}. 
In all these approaches, the R-matrix, which is the quantum group structure constant, is introduced by hand in the discretization and quantization processes. 

There have also been many attempts to try to understand the appearance of quantum groups from a physical perspective. In \cite{Freidel:1998pt}, it was argued that quantum group deformation perturbatively appears in the limit of small cosmological constant.
The works \cite{Noui:2011aa, Pranzetti:2014xva} showed that the quantum group structure could appear in the regularization of the Hamiltonian constraint. 
In  \cite{Bonzom:2014bua} a deformation of the Hamiltonian constraint, such that its kernel contains the TV amplitude, was found. 
We should also mention  
 the seminal works  \cite{Gawedzki:1990jc, Falceto_1993}, where the quantum group symmetry is identified at the classical level for the Wess-Zumino model. While this is not the gravity context, the approach used there was an inspiration for our current work.

Despite all these attempts, no actual derivation of the TV model from a gravity action exists. 
Not to the level of satisfaction achieved for the Ponzano-Regge model where undeformed symmetry appears \cite{Freidel2004, Barrett:2008wh, Freidel:2002dw, Bonzom:2012mb}. 
All the justifications listed here point to the fact that the quantum group is the right symmetry to implement in the discrete and quantum regime, 
and that this symmetry somehow respects the dynamics of the theory.
However, it is unclear what this symmetry exactly corresponds to. It cannot merely be gauge invariance since the Lorentz gauge group is independent of the cosmological constant. Also, it has to be appreciated that quantum groups introduce a preferred direction that selects a Cartan subalgebra from the onset. The source and nature of this preferred direction have been a long-standing puzzle.

\medskip

The question we would like to address here is \textbf{what is the classical origin of these quantum deformed symmetries, \textit{starting from the gravity action}?} 

\medskip

Answering this question relies on three concepts. 
The first  key  idea was first formulated in \cite{Freidel:2015gpa}, further  formalized in \cite{Donnelly:2016auv} and developed in  \cite{Freidel:2016bxd, Freidel:2019ees} at the quantum level.
Concretely, these works  establish that there are, actual symmetries in gravity represented by non-trivial canonical generators.  These symmetries reveal themselves once we 
 decompose a gravitational system into subsystems. Then the boundary of the subsystem decomposition supports the symmetry generators.
The point is that these boundary symmetry generators are the relevant symmetry generators   that one needs to use in order to construct the quantum theory.
The quantum spacetime is then obtained as a fusion of quantum representations of the boundary symmetry group. This represents the quantum equivalent of the gluing of subregions.
This idea  is built upon the works of many who have demonstrated the central importance of boundary symmetry algebra in  gravity \cite{Regge:1974zd, Iyer:1994ys, Balachandran:1994up, Balachandran:1995qa, Wald:1999wa, Carlip:1999cy, Szabados:2005wi } and developped the understanding of the nature of entanglement entropy in gauge theory
\cite{Balachandran:1995iq, Buividovich:2008yv, Donnelly:2011hn, Casini:2013rba, Radicevic:2014kqa,Donnelly:2015hxa,Lin:2018bud}.

The second and related idea, first proposed in \cite{Freidel:2011ue}, is that one can think of the process of discretizing a field theory, while respecting the
 bulk gauge invariance \cite{Freidel:2002dw, Dittrich:2012qb} as a  two-step process. The first step, that we just discussed,  is the decomposition of the system into subregions  and the second step is a coarse-graining operation where one replaces each cell of the decomposition by a 
 vacuum solution of the bulk constraints. Consequently, 
the subregion boundaries, and their symmetry charges,  encode all the relevant degrees of freedom of this corase grained data. This procedure leads to a discretization that respects, by construction, the fundamental invariance of the theory under study. It also leads to a new way to approach the continuum limit as a condensation of charge defects \cite{Delcamp:2016yix}. The choice of a solution on each cell corresponds to a vacuum choice at the quantum level \cite{Dittrich:2016typ}.
 This strategy has been developed in the case of three-dimensional gravity in \cite{Dupuis:2017otn, Freidel:2018pbr, Shoshany:2019ymo}.
 
 The third concept is illustrated in the section II for 3d gravity and in  \cite{Edge-Mode-II}  for 4d gravity. It uses the fact that it is possible to modify the expression of the boundary symmetries and their charges by the addition of boundary terms to the action.  In the case of 3d gravity, the boundary symmetry is composed of 
 the internal Lorentz symmetry and the translation symmetry.
 We show that it is necessary, in the presence of a non-vanishing cosmological constant, to add a boundary term to the action to ensure that the boundary translational symmetry is closed as an algebra. This boundary term, which implements a canonical transformation in the bulk, is the continuum analog of the classical R-matrix. It is given for 3d gravity by
 \be
 \int_{\partial M} \mr_{ij}e^i\wedge e^j, \qquad \mr_{ij}\equiv \epsilon_{ijk}n^k, 
 \ee
 where $n^k$ is a fiducial vector that is shown to be the quantum group preferred direction
 and whose norm square is proportional to the cosmological constant.
 We show that the presence of this boundary term affect the  bulk connection 
 and deforms the notion of gauge invariance, by replacing the usual gauge invariance by an equivalent one preserving the fiducial vector $n^i$.
 The fact that this is possible to introduce a fiducial vector without breaking, only deforming, gauge invariance is the central physical mechanism behind the appearance of quantum groups.
 It happens because the vector labels a bulk canonical transformation whose rotation can be rectified by a canonical boundary transformation.
 It is well-known that the charges of local rotations are given by the  boundary coframe, that they form an algebra denoted $\su$ and that the charges of local translations are given by the  boundary connection 
 \cite{Geiller:2017whh}. After deformation we find that the translation generators form a subalgebra denoted $\an$:
 \be
 \{\tP_\alpha,\tP_\beta\} ={\tP}_{ (\alpha\times \beta) \times n }, \qquad {P}'_\phi = \oint_{\pa\Sigma} \phi^I \om_I,
 \ee 
 where $\times$ denote the cross product,  $\Sigma$ is a 2d subregion and $\omega$ the (deformed) gravity connection.
 We also find that the cosmological constant enters, through $n$, in a deformation of the Lorentzian Gauss law. This gives us our first hint of the presence of a quantum group in the continuum theory.
 
 \medskip
 
 In section III, we study the process of subdivision and coarse-graining as described previously. We show that after a choice of vacuum solution on each cell, the symplectic form of the continuum theory becomes finite-dimensional. It decomposes as a sum over 
 the intersections of  cells, these are the ``links'' of the decomposition.
 
 For each link $\ell$ (and its dual $\ell^*$), we identify two holonomies $(H_\ell,\tilde{H}_{\ell})$ belonging to the rotation group $ \SU$ and two holonomies $(L_{\ell^*},\tilde{L}_{\ell^*})$ belonging to the group $\AN$ and we show that they form a \emph{ribbon structure}:
 \be
\tilde H_\ell \tilde L_{\ell^*} = {L}_{\ell^*} {H}_{\ell}. 
 \ee 
 The crux of the paper consists in proving that the phase space attached to each link is in fact the 
 \emph{Heisenberg double} $\DS$. 
 As a manifold, the Heisenberg double  is the cross-product group 
$ \DS=\SU \bowtie \AN$ defined by the ribbon structure. 
The Poisson bracket we derived is compatible with the action of a  Poisson-Lie group, which is the classical analog of the quantum group.
The fact that classical analog of quantum group symmetries  appears naturally 
when the phase space is a Heisenberg double has been established for a long time  \cite{SemenovTianShansky:1985my, SemenovTianShansky:1993ws, Alekseev_1994}.

Note that in \cite{Bonzom:2014wva}, a discrete model based on  Heisenberg doubles attached to links was proposed. It was also argued there that it provides a discretization of 3d gravity with a non-zero cosmological constant, and later on, it was shown to lead to the Turaev-Viro amplitude upon quantization \cite{Bonzom:2014bua}. The relation with the classical continuum variables was missing. The derivation of this structure from the continuum action constitutes the main result of our work.

\medskip

The article is organized as follow. 
Section I is essentially a review of existing material.  We first recall the Hamiltonian analysis of 3d gravity with a non-zero cosmological constant. We emphasize that the rotational symmetry does not depend on the cosmological constant, so that it is not clear at first why a deformation of the symmetries should appear upon quantization. 


In section II, we introduce  the relevant boundary action which   provides the right starting point for the discretization. We perform the Hamiltonian analysis of the  action in these new variables. In particular, we obtain  new rotational symmetries which do depend on the cosmological constant. 

Section III provides the main result of the paper. We provide a detailed proof that the Heisenberg double phase space is obtained from our discretization.  We highlight how the discretized variables we have obtained are related to the ones introduced in \cite{Bonzom:2014wva}. We show explicitly how the deformed symmetries of the Heisenberg double are recovered.   

In Section IV, we recall how the quantum group structure appears from the quantization of the  discrete variables we have constructed, following \cite{Bonzom:2014bua}.

\section{Canonical analysis of the 3d gravity action with a cosmological constant} \label{normal}

We first recall the standard canonical analysis of the first order 3d gravity action with a non-zero cosmological constant. We consider a 3-dimensional   manifold $M$ \cite{Carlip:1998uc}. The greek indices $\alpha,\beta,..\in\{1,2,3\}$ are spacetime indices, while capital latin letters $I,J,..\in\{1,2,3\}$  are internal indices.

\paragraph{From metric formulation to first order formulation.}
 In the metric formulation the action is given by 
\begin{align}
S_{EH}[g_{\mu \nu}]&=- \frac1{2\sigma \kappa} \int_M{d^3x \,\sqrt{\sigma\det(g_{\mu \nu})} \: \left(R[g_{\mu \nu}]  - 2 \Lambda\right)}\label{eq:903},
\end{align}
where $\kappa={8\pi G}$  and $\sigma$ encodes the signature, $\sigma=-1$ for the Lorentzian case and  $\sigma=+1$ for the Euclidean case. We introduce the frame field  $e^I_{\mu}$, such that 
\begin{equation}
g_{\mu \nu} = \eta_{IJ} e^I_{\mu} e^J_{\nu} \quad , \qquad e^I_{\mu} e^{\mu}_J = \delta^I_J , \quad e^I_{\mu} e^{\nu}_I = \delta^{\nu}_{\mu}\,.
\label{eq:13}
\end{equation}
The internal metric is then $\eta = (+,+, \sigma)$. We also introduce the spin connection $\tilde A_{IJ}$, a $\so(\eta)$ valued spin connection, such that $\tilde A^{IJ}=-\tilde A^{JI}$. The associated curvature is 
\be
R^{IJ}[\tilde{A}] = \text{d} \tilde{A}^{IJ} + \tilde{A}^{I}_{\: \: \, L} \wedge \tilde{A}^{LJ}. 
\ee
Replacing these quantities in the action \eqref{eq:903}, we recover
\begin{equation}
S_{GR}[\tilde{A},e] = -\frac1{2\sigma \kappa}  \int_M{\varepsilon_{IJK} \left(e^I \wedge R[\tilde{A}]^{JK} - \frac{\Lambda}{3} \, e^I \wedge e^J \wedge e^K\right)}\,.
\label{eq:bf2}
\end{equation}
This is the \textit{first order formulation} of three dimensional gravity. When starting from this action, the frame field and the connection can be taken as independent variables. We note that we do not have to assume that the frame field is invertible.  As such, the solution space of the first order formulation is larger than the second order one. This is a standard approach since the seminal work by Witten on the 3d gravity Chern-Simons formulation \cite{Witten:1988hc}.  

In the action \eqref{eq:bf2}, it is common to rewrite the connection with a single index, using the Levi-Civita tensor, which also depends on the signature.  Fixing $\epsilon_{123}=1$, we have  $\epsilon^{123}=\sigma$ and furthermore 
\begin{align}
\epsilon^{\mu \nu \rho} \epsilon_{\mu \beta \gamma} &=  \sigma (\delta^{\nu}_{\beta} \delta^{\rho}_{\gamma}-\delta^{\rho}_{\beta} \delta^{\nu}_{\gamma}  ) \,.\label{eq:neu4}
\end{align}
We have then 
\bes
&& \tilde{A}^J=   \frac{1}{2} \, \epsilon^J_{\: \: KL} \tilde{A}^{KL}, \quad \tilde{A}^{JK}=\sigma \epsilon^{JK}{}_I\tilde{A}^I \\
&&R^J = \frac{1}{2} \, \epsilon^J_{\: \: KL} \, R^{KL}, \quad R^{JK}=\sigma \epsilon^{JK}{}_IR^I, \quad R^I= \rd \tilde{A}^I -\frac{\sigma}2\epsilon^I{}_{JK}\tilde{A}^J\wedge \tilde{A}^K. 
\ees
In order to have a curvature formula that does not depend on the signature, we can rescale the connection $A=-\sigma \tilde{A}$, so that 
\be
R^I[\tilde{A}]=-\sigma F^I[A] = -\sigma (\rd A^I + \demi \epsilon^{I}{}_{JK} A^J\wedge A^K).
\ee
$ A$ is still a $\so(\eta)$ valued spin connection.   Replacing this expression in the action, we obtain 
\bes
S_{BF}[A,e] &=&  \frac1\kappa \int_M{ \left(\eta_{IJ} e^I \wedge F^{J}[A] +\sigma \frac{\Lambda}{6}\epsilon_{IJK}  \, e^I \wedge e^J \wedge e^K\right)} \label{action}\\
&=& 
\frac1\kappa  \int_M \,   e^I \wedge  \left(F_I[A] + \sigma\frac{\Lambda}{3} \,E_I \right)
\,, 
\ees
where  $E_I =\demi \,   (e \times e)_I $ is the area flux,   $F_I[A] 
\equiv \text{d}A_I +  \demi \,  (A \times A)_I$ denotes the curvature of $A$ and $(A\times B)^I  = \epsilon^{I}_{JK} A^J\wedge B^K$ denotes the cross-product of Lie algebra valued forms.  In the following, we will work in units where $\kappa=1$, re-establishing the units when deemed useful.

%
%

\medskip

\paragraph{Equations of motion.} 
One can couple this action to matter field via $S^{\mathrm{Mat}}(e,A;\phi)$ and we denote $\Pp_I \equiv-\frac{\delta S^{\mathrm{Mat}} }{\delta e^I}$ the energy momentum density and 
$\Sp^I \equiv-\frac{\delta S^{\mathrm{Mat}} }{\delta A_I}$ the angular-momentum density of the matter fields.  The equations of motion are given by 
\bes
 F_I[A] + {\Lambda} \,\sigma E_I \approx \Pp_I \qquad  \rd_A e ^I \approx \Sp^I, 
\ees
where $\rd_A e^I \equiv   \rd e ^I + (A\times e)^I$ is the torsion of $A$.
In vaccuum, when no matter is present,  the first equation is the curvature constraint $F_\Lambda^I\equiv F[A]^I + {\Lambda} \, E^I \approx 0$ and the second equation is the torsion free condition since $T^I\equiv \rD e^I\approx 0$. 
We use the notation $\approx$ to stress that we have implemented the equations of motion.

\medskip

\paragraph{Action symmetries}
The action is invariant under a set of (gauge) symmetries. The first obvious symmetry is given by the  $\so(\eta)$ infinitesimal gauge transformations, parametrized by the scalar fields  $\alpha^J$, 
\bes \label{SU2transfo}
\delta_{\alpha}e^I&= & (e\times \alpha)^I,
\qquad 
\delta_{\alpha}A^I = \rD\alpha^I,
\\  \delta_{\alpha}\Sp^I&=& (\Sp \times \alpha)^I,\qquad  
 \delta_{\alpha}\Pp^I= (\Pp\times \alpha)^I.\nn
\ees
They do not depend on the cosmological constant.
\\
The second one is the "shift" symmetry, parametrized by  the scalar fields  $\phi^J$,   
\bes \label{shift}
\delta_{\phi}e^I&= & \rD\phi^I,  \qquad \qquad
\delta_{\phi}A^I= \Lambda\, (e \times  \phi)^I,  \\
\delta_{\phi}\Sp^I &= &  (\Pp \times \phi)^I, \qquad 
 \delta_{\phi}\Pp^I=  \Lambda\,  (\Sp\times \phi)^I.\nn
 \label{transsym2}
\ees
These transformations are $\Lambda$ dependent. 
The last identity means that in the presence of a non-zero $\Lambda$, the notion of energy and momentum depends on the translational frame via the angular momenta density. In the same way that the notion of angular momenta depends on the rotational frame via the energy momentum density.
\\
Diffeomorphism symmetry can be written, on-shell of the equations of motion, as a combined action of gauge and shift symmetries with field dependent parameters    \cite{Horowitz:1989ng}. Given an infinitesimal diffeomorphism $\xi$, we define the field dependent parameters
\be
\alpha_\xi ^I = \iota_\xi A^I,\quad \phi_\xi^I = \iota_\xi e ^I,
\ee
and we can express the action of an infinitesimal diffeomorphism as a gauge or shift symmetry on-shell 
(in the vacuum case). 
\bes
 \pounds_\xi A^I & =& \rd \iota_\xi A^I +   \iota_\xi \rd A^I = \iota_\xi F^I_\Lambda+\delta_{\alpha_\xi}A^I + \delta_{\phi_\xi}A^I  \approx \delta_{\alpha_\xi}A ^I+ \delta_{\phi_\xi}A^I \\ 
 \pounds_\xi e^I &=&\rd \iota_\xi e^I +   \iota_\xi \rd e^I =  \iota_\xi T^I  +\delta_{\alpha_\xi}e^I + \delta_{\phi_\xi}e ^I\approx  \delta_{\alpha_\xi}e^I + \delta_{\phi_\xi}e^I.
\ees

\medskip

\paragraph{Symplectic form and Poisson brackets.} 
Let us now perform the Hamiltonian analysis of the action \eqref{action}. We consider $M = \R \times \Sigma$. The symplectic potential associated with $S^{\text{BF}}_M$ is 
identify as the boundary variation $\delta S^{\text{BF}}_M \approx \Theta^{\text{BF}}_{\pa M}$. 
The symplectic form $\Omega^{\text{BF}}_\Sigma = \delta \Theta^{\text{BF}}_\Sigma$, associated with a Cauchy slice $\Sigma$ is 
\be \label{form0}
\Theta^{\text{BF}}_\Sigma = -\int_{\Sigma} \la  e\wedge \delta A \ra, \quad \Omega^{\text{BF}}_\Sigma  = -\int_{\Sigma} \la \delta e\curlywedge \delta A \ra ,
\ee
where $\delta$ encodes the field variations,
$\curlywedge$ is the extension of the wedge product to variational forms\footnote{ More specifically, we have a bi-complex structure: we have a differential $\rd$ on space(-time) and a differential $\delta$ on  field space. If $\alpha$ is a degree $a$ form and $\beta $ a degree $b$ form \textit{in space}, we have 
\be
\alpha \curlywedge \beta :=\alpha\wedge \beta,
\qquad
\alpha \curlywedge \delta \beta := \alpha \wedge \delta \beta,\qquad
\delta \alpha \curlywedge \delta \beta := - (-1)^{ab} \delta \beta \curlywedge \delta \alpha.
\ee },
 and the pairing is given by $
\la \delta e \curlywedge \delta A \ra=\eta_{IJ} \delta e^I \curlywedge \delta A^J$. 
Accordingly, the canonical variables are the pairs 
$(A^I_a(x),  e_{b}^J(x))$ where $a,b$ are indices tangent to $\Sigma$, $a,b,..\in\{1,2\}$. The canonical Poisson bracket generated by \eqref{form0} is simply, $\forall x,y\in \Sigma$,
\beq\label{symplectic 0}
\poi{A^I_a(x), e_{b}^J(y)}= \kappa\, \epsilon_{ab}\, \eta^{IJ} \,\delta^2(x-y), \quad \poi{A^I_a(x), A_b^J(y)}=0=\poi{e_{a}^I(x), e_{b}^J(y)},
\eeq
 where we reinstated $\kappa$ for  completeness. 

\medskip
\paragraph{Charges algebra}
It is well-known that the total Hamiltonian and the generators of rotational and translational symmetry are given by boundary terms and satisfy a closed algebra.
Let us recall that the Hamiltonian generator associated with a canonical field transformation  
$\delta_\psi$ is  $H_\psi$   provided we have 
\be\label{symp}
\delta_\psi \ip \Omega = \int_{\Sigma}( \la \delta_\psi e\curlywedge \delta A \ra-
 \la \delta e\curlywedge \delta_\psi A \ra) =-\delta H_\psi. 
\ee 
The Poisson bracket of two generators is defined to be
\be\label{Poisson}
\{H_\psi,H_{\psi'}\} = \Omega(\delta_\psi,\delta_{\psi'}) =\delta_\psi H_{\psi'}. 
\ee
In other words, the condition \eqref{symp} means that the
Hamiltonian generator $H_\psi$ generates the canonical transformation 
\be
\delta_\psi \cdot  =\{H_\psi,\cdot\}.
\ee
One denotes $J_{\alpha}$ the generator of rotational symmetry
($\delta_\alpha =\{J_\alpha,\cdot\}$) , $P_\phi$ the generator of translational symmetry. 
They are given by 
\bea\label{generator 1}
J_\alpha &=& \int_{\Sigma}  \alpha_I (\Sp^I - \rd_Ae^I)+\oint_{\pa\Sigma } \alpha_I e^I,\cr
 P_\phi &=&\int_\Sigma \phi^I \left(\Pp_I - F_I(A)- \sigma\tfrac{\Lambda}2 \, (e \times e)_I\right) +\oint_{\pa\Sigma} \phi^I A_I,
\eea
 The transformations associated to a parameter vanishing on the boundary  are gauge transformations. Hence they have a vanishing charge. Their canonical generator vanishes  on-shell since it is  proportional to the constraints. On the other hand, transformations whose boundary parameters do not vanish, have non vanishing charges. They are the boundary symmetries.  The corresponding boundary charges are given by 
\be
 J_\alpha
 \approx \oint_{\pa\Sigma } \alpha_I e^I,\qquad
 P_\phi \approx \oint_{\pa\Sigma} \phi^I A_I. 
\ee 
Using \eqref{Poisson} and the expressions (\ref{SU2transfo},\ref{shift}) for the transformations, one can evaluate the boundary charge algebra  (reinstating $\kappa$)
\bes 
&&\{J_\alpha,J_\beta\}= \ka \, J_{(\alpha\times \beta)},\quad
\{P_\phi,P_\psi\} =  { \sigma}\, \ka\, \Lambda \,  J_{(\phi\times \psi)},\nn\\
&&\{J_\alpha, P_\phi \} =\ka \, P_{(\alpha\times \phi)} + \ka \oint_{\pa\Sigma} \phi^I \rd\alpha_I. \label{boobost}
\ees
One sees that there exists a central extension in the commutator between $J_\alpha$ and $P_\phi$. 
Therefore this algebra is first class only for the transformation parameter $\alpha$ that is  constant on $\pa \Sigma$.


In the following, we will be interested in the transformations that are global and hence such that both $\alpha$ and $\phi$ are constant. In this case, the charges   form a finite dimensional Poisson Lie algebra. 

\medskip

\paragraph{Quantum algebra of observables.} 
The corresponding quantum operators  for the global charges are given by
\be
 \hJ^I
 \approx i\oint_{\pa\Sigma }\hat e^I,\qquad
 \hP^I \approx i\oint_{\pa\Sigma} \hat A^I.  
\ee 
 We require them to be be  antihermitian $\hJ^\dagger=-\hJ$, $\hP^\dagger=-\hP$. They satisfy the Lie algebra brackets 
\beq\label{algebra}
[\hJ_I,\hJ_J]= l_P \,\epsilon_{IJK}\, \hJ^K, \quad [\hJ_I,\hP_J]=l_P\, \epsilon_{IJK}\,\hP^K, \quad [\hP_I,\hP_J]=\sigma \,l_P\,  \Lambda \,\epsilon_{IJK}\,\hJ^K, 
\eeq 
with $l_P=\hbar\ka$ the Planck length.  The indices are raised with the  metric $\eta_{IJ}$. 
Hence according to the signature $\sigma$ and the sign $s$ of the cosmological constant $\Lambda$, the quantum algebra of charges is isomorphic to a  well-known Lie algebra  $\ds_{\sigma s}$.  We have  
$\ds_{++}=\so(4)$ when dealing with a spherical space-time $S_3$,   $\ds_{+-}=\sl(2,\C)= \ds_{-+}$ when dealing with a hyperbolical space-time $H_3$ or with a de Sitter space-time $dS_3$ 
and finally $\ds_{--}=\so(2,2)$ when dealing with an anti de Sitter space-time $AdS_3$.

\medskip

 \paragraph{Gauge theory for $\ds_{\sigma s}$.} Let us note the generators of Lie algebra $\ds_{\sigma s}$ by $\bfJ_I$ and $\bfP_J$, respectively the Lorentz/rotation generators and the boosts. 
To build the action, we introduce a pairing between the generators, i.e. an invariant bilinear form over $\ds_{\sigma s}$. The relevant one  is\footnote{See \cite{Meusburger:2007ad} for a discussion on the most general pairing one can consider.}  , 
\be\label{pairing}
\la \bfJ_I,\bfP_J\ra = \eta_{IJ} = \la \bfP_I,\bfJ_J\ra , \quad 
\la \bfJ_I,\bfJ_J\ra = 0 = \la \bfP_I,\bfP_J\ra.
\ee
The frame field has value in the boosts, $e\equiv e^I\bfP_I$, whereas the connection has value in $\so(\eta)$, $A\equiv A^I \bfJ_I$. Hence, the curvature $F[A]$ is an object with value in $\so(\eta)$, whereas the torsion $T[e,A]$ takes value in the boosts. In particular the covariant derivatives can be expressed in terms of the structure constant of $\so(\eta)$. 
\bes
\rD \alpha = \rd \alpha + [A,\alpha], \quad \textrm{ with }  \alpha= \alpha^I\bfJ_I, \nn \\
 \quad \rD \phi = \rd \phi + [A,\phi], \quad  \textrm{ with } \phi= \phi^I\bfP_I.
\ees

\medskip
We could now try to construct the LQG kinematical Hilbert space by imposing the Gauss constraint first as usually done. 
Since the rotational charge  does not depend on $\Lambda$, we expect to recover after discretization the standard spin networks based on $\SU(2)$, just as when $\Lambda=0$. Hence the kinematical states are not given in terms of a quantum group structure. 
However we know that the quantum group structure needs to appear once we properly implement the dynamics. For example in the Turaev-Viro  model \cite{TV}, which gives the proper quantization of 3d gravity,  the boundary states are given in terms of quantum group spin networks.
This raises a fundamental puzzle and shows that the choice of discretization scheme could be at odd with the dynamics of the theory.  
While both formulations (with group or quantum group  spin networks) should agree in the continuum limit, it is not clear how to define  the quantum theory with undeformed spin networks and then to achieve a proper continuum limit, while the Turaev-Viro model is well-defined and also known to be invariant under refinement  therefore defining a continuum theory. Resolving this tension means that   one needs to deal at the classical level with a different rotational charge, which should depend on $\Lambda$. 

Note also that an essential step to construct the quantum states is to discretize the theory, and in particular the charge information. We note that the translational charge algebra \eqref{boobost} does not form a closed algebra, rendering its discretization more obscure.  
As we will show modifying the  rotational charge in a $\Lambda$ dependent way allows to perform the discretization without breaking the symmetry.

%
%
%

\section{New variables and new action} \label{sec:new action}
In order to change the rotational charge structure, which should also depend on $\Lambda$, it is natural to add a boundary term. 

\subsection{Gravity Action and canonical transformation}\label{sec:canon}
 
\paragraph{Boundary term and canonical transformation.}  Let us consider   a general vector $n^I$ parametrizing the boundary contribution. 
 We will see what further conditions $n$ is  required to satisfy along the way.
We consider then the original action \eqref{action} modified by the boundary term\footnote{QG stands for quantum group.}
\bes \label{rmatbdy}
{S}_{QG}[e,A]&\equiv
& 
S_{BF}[e,A]+ \frac1{2\ka} \int _{\partial M}\, \mr_{IJ}\,  e^I\wedge e^J\\
&=& S_{BF}[e,A]+
\frac1{2\ka} \int_{\partial M}  (e\times e)_I n^I \nn\\
 &=& \frac1\ka\int_M \,   e^I \wedge  \left(F_I[A] + \sigma\frac{\Lambda}{6} \, \epsilon_{IJK} \,   e^J \wedge e^K \right) + \frac1{2\ka} \int_{ M} \rd \left( (e\times e)_I n^I\right). \label{action1}
\ees
The boundary term does not modify the equations of motion. We note that while $n$ is defined first on the boundary $\partial M$, it can be naturally extended to the bulk $M$ using Stokes theorem. As before we will work with $\ka=1$ until deemed necessary. 

\medskip

To perform  the Hamiltonian analysis of the new action, we assume  as before that $ M=\R\times \Sigma$. The new symplectic potential is  
 \be\label{new theta}
 \Theta_{QG} = \int_{\Sigma} e _I\wedge \delta A^I 
 -\frac12 \delta \int_{\Sigma}(e\times e)_I n^I =  \int_{\Sigma} e_I\wedge \delta \tom^I -\demi \int_{\Sigma} (e\times e)_I \cdot \delta n^I,
 \ee
 where we have introduced a new connection
 \be
 \omega^I\equiv A^I +(n\times e)^I. 
 \ee
%
%
%
We see from \eqref{new theta} that we have   an extra  pair of conjugated variables $(n, E= \demi (e\times e))$ where the area flux $E$ is conjugated to $n$. We note that if $n$ is treated as a kinematical structure, it is required to be \textit{constant as a field}, $\delta n=0$, and  the boundary term simply induces  a canonical transformation (\textit{in the bulk}) that  modifies the original symplectic potential \eqref{form0}.
Note that this conditions forbids the vector $n^I$ to be related to the boundary normal\footnote{ If we denote $ s_a $ the normal form to the boundary, we can construct, using the frame, the 
internal normal $ s_I = e_I^a s_a$. This normal is field dependent $\delta s_I = \delta e_I^a s_a\neq 0$, where we use that the boundary normal form is field independent: $\delta s_a=0$.
Therefore the vector $n^I$ being kinematical cannot be related to the boundary normal.}.

 This canonical transformation only modifies the connection.   \textit{We will assume that $\delta n=0$  from now on}.
Hence $(e_a^I,\omega_b^J)$ is our new canonical pair, $\forall (x,y)\in\R^2$,
 \beq\label{new poisson}
\poi{\om^I_a(x), e_{b}^J(y)}= \ka \, \epsilon_{ab}\,  \eta^{IJ} \, \delta^2(x-y), \quad \poi{\om^I_a(x), \om_b^J(y)}=0=\poi{e_{a}^I(x), e_{b}^J(y)},
\eeq

With such a change of variables, we can express the  curvature in terms of the new connection $\omega$ 
\bes \label{F(A)}
&&F[A]= F[\tom + e\times n] = F[\tom]+ \rd_\tom (e\times n) + \demi (e\times n)\times (e\times n), 
\ees
where $\rd_\omega\alpha =\rd \alpha +\omega\times \alpha$. To evaluate the action in terms of $\omega$, one establishes\footnote{This follows from 
\be
\frac{n^2}3 e\cdot (e\times e)= (e\cdot n) (n\cdot (e\times e)),
\ee
and the cross-product identity 
$
(\alpha\times \beta)\times \gamma = \sigma[ (\alpha\cdot \gamma)\beta -\alpha (\gamma\cdot \beta) ].$} that 
\bes 
 \demi e\cdot ((e\times n)\times (e\times n))=\frac{\sigma n^2}{6}  e\cdot (e\times e).
\ees

We choose the  normalization  $n^2=- \Lambda$, 
 as a new restriction on $n$, so that the last term of \eqref{F(A)}   compensates the term proportional to $\Lambda$ in the action \eqref{action1}. 

With the assumptions that $\delta n=0, $ and $n^2=- \Lambda$, the change of variables implies that the  action \eqref{action1} becomes
\bes \label{QGp}
 {S}_{QG}= \int_M \,   \left( e\cdot F [A] + \sigma\frac{\Lambda}{3} \,e\cdot E \right) +  \int_{\partial M} E  \cdot n 
&=& \int_M\, \left( e \cdot  F[\tom] 
-  E  \cdot \rd_\tom n  
\right).
%
\ees  
While the original action \eqref{action1} couples the frame $e$ and flux $E =\tfrac12 (e\times e)$ the modified action is achieving a ``separation of variables'' where $e$ and $E$ are decoupled. This will simplify the analysis of the theory and its symmetries.

\medskip 

  The equations of motion of the new action \eqref{QGp} are now 
\bes \label{motioneom}
 F_I[\tom] - (e\times \rd_\tom n )_I \approx  \tPp_I
\qquad \mathrm{and} \qquad &   \rd_\tom e^I + \frac12 [(e\times e )\times n]^I\approx \tSp^I .
\ees
The matter spin density $\tSp\equiv -\frac{\delta S_{\mathrm{QG}}}{\delta \omega}$ is unchanged while  the energy-momentum density 
$\tPp\equiv -\frac{\delta S_{\mathrm{QG}}}{\delta e}$ is redefined\footnote{One uses that 
$-\delta S_{QG}=\cP \delta e + \cJ \delta A = \cP'  \delta e + \cJ' \delta \omega $}:
\bes
  {\tSp}_I =  \Sp_I,\qquad {\tPp}_I =  \Pp_I + (n\times \Sp)_I. \label{newSET}
\ees


\medskip

\paragraph{Nature of the vector $n$. } In the Euclidean case $\sigma=+1$,  the normalization  condition 
$n^2=- \Lambda$ can be achieved by  a real vector in the hyperbolic case ($\Lambda<0$) or by a pure imaginary vector in the spherical case ($\Lambda>0$). If $\Lambda=0$, then either $n=0$ or it is specified by a Grassmanian number. 

In the Lorentzian case, $n$ is time-like (or \textit{imaginary} space-like) for the de Sitter  case  and space-like (or \textit{imaginary} time-like) for the AdS case. When $\Lambda =0$ we have two options,  $n $ is either a non trivial  null vector or it simply vanishes. 

Note that in either signature, dealing with a purely imaginary contribution and hence a complex action leads to some subtle discussion about reality conditions (reminiscent of adding a purely imaginary Immirzi parameter in 4d). Since this technical aspect goes beyond our main point we do not dwell on it.  

\begin{center}
\begin{tabular}{ c|c|c } 
& Euclidean & Lorentzian \\\hline
Flat: $\Lambda=0$ & $n=0$ or $n$ is Grassmanian&  $n=0$ or $n$ is light-like\\\hline
AdS:$\Lambda<0$ &$n$ is space-like& $n$ is space-like or \textit{imaginary} time-like \\\hline
dS:$\Lambda>0$ &$n$ is \textit{imaginary} & $n$ is time-like or \textit{imaginary} space-like\\ \hline
\end{tabular}
\end{center}

\medskip

\paragraph{Symmetries of the action. }

Since the action $S_{\textrm{QG}}$ depends explicitely 
on a vector $n$, one might worry that this vector acts as a background structure and that  this action explicitly breaks local rotational symmetry. It turns out, quite remarkably,  that this is \textit{not} the case. The action is still invariant under  gauge transformations generalizing the local $\SO(\eta)$ transformations \eqref{SU2transfo} and the shift transformations \eqref{shift}.

First let us notice that since we required $n$ to be constant as a field $\delta n=0$ this implies that it will not change under the symmetry transformations, spanned by the Hamiltonain generators $H_\psi$ (with $\psi=\alpha, \phi$),
\be
\delta_{\psi}n =\{ H_\psi,n\}=0.
\ee
As a consequence, $n$ can be seen as a scalar for the different gauge transformations. In the following, we are going to determine the shape of the gauge transformations on the field $e$ and $\omega$ which are consistent with this constraint $\delta_\psi n=0$.  In order to distinguish the new infinitesimal transformations from the previous one, we will note them $\delta'_\psi$. We demand therefore that $\delta'_\psi n=0$, for $\psi=\alpha, \phi$.

\smallskip

Let us  study the set of transformations, generalizing the $\so(\eta)$ infinitesimal transformations, that we parametrize by $\alpha^I$. Since we have that  
 \bes
 \delta_\alpha' n^I&=& 0,
 \ees
 and that we still have that $e^I$ should transform as a vector, 
 \bes
\delta'_{\alpha}e^I&= &  (e\times \alpha)^I =\delta_\alpha e^I.\label{transfosu1} 
\ees
We can use the transformations of $A$ and the relation between $A$ and $\omega$ to infer the transformations of $\omega$.
\bes
\delta'_\alpha A = \delta'_\alpha(\om - n\times e)= \delta'_\alpha\om - n\times \delta'_\alpha e \Leftrightarrow
\delta'_{\alpha}\tom^I= \rd_\tom  \alpha^I + (e\times (n\times \alpha))^I \equiv D \alpha^I.
\label{transfosu2}
\ees

 The second set of transformations, parametrized by $\phi$ generalizes the shift symmetry. We still demand that $\delta'_\phi n^I=0$. We have
\bes
\delta'_{\phi}\tom^I&= & (\phi\times \rd_\omega n )^I 
 \label{transfoan2}\cr
\delta'_{\phi}e^I &= &\rd_\om \phi^I  + (  (e  \times \phi)\times n )^I  \equiv \tilde{D} \phi^I \label{transfoan1}.
\ees
These transformations satisfy $\delta'_\phi e^I = \delta_\phi e^I+ \delta_{\alpha=\phi\times n}e^I$. 

\medskip

It is worth noticing that now  \textit{both types of gauge transformations  are dependent on the cosmological constant}  through the vector $n$ and  both leave the auxiliary vector $n$ invariant. We emphasize again that this implies that the vector $n$ is a \textit{scalar} for such gauge transformations. 

\medskip 
The proof of the invariance of the action follows from the (generalized) Bianchi identities. We have indeed that 
\beq
-\delta S_{QG}= (F-e\times \rd_{\omega}n)\delta e + (\rd_\omega e + [(e\times e)\times n])\delta \omega = \mF \delta e + \mT \delta \omega.
\eeq 
Plugging the symmetry transformations  \eqref{transfoan1} or \eqref{transfosu2} and  \eqref{transfosu1}, and using some integration by parts, we can use the relations 
\bes
D\mF + \mT \times (\rd_\omega  n)    =0\\
\tilde D \mT + e \times  \mF  =0, 
\ees
which generalize the notion of Bianchi identity to the case where we deal with a matched pair of Lie algebras, see Appendix \ref{big-guy} and  \cite{Girelli:2021pol}.

\subsection{Deformed boundary symmetry algebra and Manin pairs}

\paragraph{New charges algebra. } One can wonder at this stage, what have we gained by going to this more elaborate description of the same physical system?
The clear advantage of this description shows up when we look at the symmetry algebra and the  transformations of the 
spin and energy momenta densities. These transformation can be deduced from (\ref{transfosu1},\ref{transfosu2},\ref{transfoan1}) by acting on the LHS of the constraints \eqref{motioneom}.
For instance one finds that 
\be
\delta'_\alpha \tSp =  \tSp \times \alpha,\qquad 
\delta'_\phi \tPp = (\tPp \times \phi)\times n ,
\ee
which shows that the modified energy-momentum density transforms homogeneously under a local translation, unlike \eqref{transsym2}. 
The charges associated with these transformations 
$ {\delta}'_\alpha \ip \Omega = -\delta {J}'_\alpha$ and $ \delta'_\phi \ip \Omega =-\delta {P}'_\phi$,  are given by 
\bea\label{new charges}
{J}'_\alpha &=&  \int_{\Sigma} \alpha_I \left[\tSp -   \rd_\omega e - \tfrac12  ((e\times e )\times n)  \right]^I +\oint_{\pa\Sigma} \alpha_I e^I, \cr
{P}'_\phi &=& \int_{\Sigma}  \phi^I \left[\tPp - F[\tom] + (e\times \rd_\tom n )\right]_I  +\oint_{\pa\Sigma} \phi^I \omega_I.
\eea
On-shell, these charges are simply 
\be\label{newcharge}
 {J}'_\alpha  = \oint_{\pa\Sigma} \alpha_I e^I= J_\alpha , \qquad 
 {P}'_\phi = \oint_{\pa\Sigma} \phi^I \om_I =  P_{\phi}+ J_{\phi\times n}.
\ee
%
%
The charge algebra is such that $J'_\alpha$ and $P'_\phi$ generate two subalgebras given by
 \bea\label{newcom1}
\{J'_\alpha, J'_\beta\} = \ka\,  J'_{\alpha\times \beta},& &\qquad
\{\tP_\phi,\tP_\psi\} =\ka \, {\tP}_{ (\phi\times \psi) \times n }+\ka\, \oint_{\pa \Sigma}(\phi \times  \psi) \cdot  \rd n,
\eea
while the cross-commutator is given by 
\bea \label{newcom2}
\{J'_\alpha,\tP_\phi\}
&=& \ka\, \tP_{ \alpha\times \phi   }  + J'_{\phi\times (\alpha \times n ) } 
+\ka\,  \oint_{\pa \Sigma} \phi\cdot \rd \alpha.
\eea
The proof is detailed in the appendix \ref{proof1}. We emphasize that we are using the simple derivative $\rd$ since $n$ is a scalar in terms of the gauge transformations. 

\medskip 
We see that the commutator of energy-momentum charges possesses a central charge if $n$ is not constant. \textit{From now on}, we assume that
$\rd n=0$. 
In  this case, we see that the modified energy momentum charges $\tP_\alpha$ form \textit{a closed subalgebra} 
and the central charge is concentrated of the bracket between rotation generators $J'$ 
translation/boost generators $\tP$. This is in sharp contrast with the original description \eqref{boobost}, where the momentum generators do not form a closed subalgebra and it is the main reason behind the canonical transformation and the normalisation $n^2 = - \Lambda$.

\medskip

\paragraph{Another condition on $n$. }
{ Before discussing the shape of the global symmetries, it will be useful to  fix for once and for all the  vector $n$.  Without loss of generality, we can always choose the vector $n$ as defining the direction $3$, 
 $n^I=(0,0,n^3)$.  As we have seen earlier in \eqref{QGp}, according to the normalization condition $n^2=- \Lambda$, the vector $n$ can be space-like or time-like, or even imaginary
. 
 Since we have fixed the direction of $n$, this means that the metric should also depend on $s$, the sign of $\Lambda$. 
Let us review the different cases.

If we are in the Euclidean case with $\Lambda>0$, then $n^I=(0,0,i\sqrt\Lambda)$ and the Euclidean metric is consistent. 
In the other cases where $\Lambda\neq0$, we will take  $n^I=(0,0, -s\sigma \sqrt{|\Lambda|})$ and a metric $\eta^{\sigma s}$ such that 
\be
\eta^{\sigma s}_{IJ}=\mathrm{diag}(+,-s\sigma,-s), \quad  n^I \eta_{IJ} n^J= -\, s \, |\Lambda|=  -\, \Lambda.
\ee
Finally, in the case where $\Lambda=0$, we  stick to the usual metric $\eta_{IJ}=\mathrm{diag}(+,+,\sigma)$. 
Fixing such convention will allow to connect more easily to the usual quantum group formalism where it is \textit{always} the third direction that is picked out as preferred.  Let us review  the full set of constraints we have on $n$,
\be\label{constraints on n}
\delta n =0 \,\, (\Rightarrow \delta'_\alpha n=\delta'_\phi n=0), \quad n^2=-\Lambda, \quad \rd n=0, \quad n^I= (0,0,n^3).
\ee

\medskip

While the symmetry structure of the metric is still isomorphic to $\so(\eta)$, the time direction is not always the same in the Lorentzian case, to account for $n$ being space-like or time-like. Let us review the different explicit forms of $\so(\eta)$. We  note $\bfJ^I$ their generators. The commutation relations are simply $[\bfJ^I,\bfJ^J] = \epsilon^{IJ}{}_K \bfJ^K$, where we use the  the metric $\eta_{\sigma s} $ to lower the index $K$.
This means that we have different algebras for different choices of $(\sigma,s)$. 
We denote the different cases by $\su_{\sigma s}$
\bes \label{generalcom}
\su_{+-}= \su(2):&\quad &
[\bfJ_1,\bfJ_2]=\bfJ_3,\quad
[\bfJ_2,\bfJ_3]=\bfJ_1,\quad
[\bfJ_3,\bfJ_1]=\bfJ_2, \nn\\
\su_{-+}=\su(1,1):&\quad &
[\bfJ_1,\bfJ_2]=-\bfJ_3,\quad
[\bfJ_2,\bfJ_3]=\bfJ_1,\quad
[\bfJ_3,\bfJ_1]=\bfJ_2 \nn\\
\su_{--}= \sl(2,\R):&\quad&
[\bfJ_1,\bfJ_2]=\bfJ_3,\quad
[\bfJ_2,\bfJ_3]=\bfJ_1,\quad
[\bfJ_3,\bfJ_1]=-\bfJ_2.
\ees
}

\medskip

\paragraph{Quantum algebra of observables.}
The   algebra given in \eqref{newcom1} and \eqref{newcom2} is  first class only for the transformation parameters that are constant  on the boundary. 
Such a set of constant parameters generates then global symmetry transformations which  form a finite dimensional Poisson Lie algebra. 
%
%
 The associated quantum algebra  is now generated  by the quantisation of the global charges
\be\label{chargedef}
\hJ'^I= \hJ^I= i \oint \hat{e}^I,\qquad \hP'_I = i \oint \hat{\om}_I. 
\ee
As we have seen in \eqref{newcharge}, we have just performed a linear change of basis, hence the global charges still form  an algebra isomorphic to $\ds_{\sigma s}$, with    $\ds_{++}=\so(4)\sim\su(2)\oplus \su(2),\,
\ds_{+-} = \sl(2,\C) = \ds_{-+},$ and $
\ds_{--}= \so(2,2)\sim \sl(2,\R)\oplus \sl(2,\R).$
The physical reality condition that arises from the quantisation of the 
global algebra with $e,\om$ real \eqref{chargedef} demands 
that all generators are antihermitian
and that the vector $n$ is real:
\be \label{real}
\hJ'^\dagger =-\hJ', \quad \hP'^\dagger= -\hP',\quad  \bar{n} =n.
\ee
We note that the Euclidean case with positive cosmological constant does not have a real $n$, hence we will not discuss it here. It requires a more careful analysis on the reality condition.

\paragraph{$\ds_{\sigma s}$ as a Manin pair.}  The  Lie algebra $\ds_{\sigma s}$ has the structure of a Manin triple, that is, it is a  classical Drinfeld double that can be written as a matching pair 
$\ds_{\sigma s}=\cs\bowtie \cs^*$ \cite{Chari:1994pz}. 
By construction, $\ds_{\sigma s}$  possesses an invariant symmetric pairing denoted $\la \cdot,\cdot\ra$ of signature $(3,3)$ and it can be decomposed as a pair of isotropic algebras
\be
\ds=\cs \oplus \cs^*,\qquad   
\la \cdot,\cdot\ra|_{\cs}=0=\la \cdot,\cdot \ra|_{\cs^*}.
\ee
The symmetric pairing is simply the canonical pairing between $\cs$ and $\cs^*$.
Given $\ds_{\sigma s}$, its  subalgebra $\cs$ is the  subalgebra $\su_{\sigma s} $
 with generators $\bfJ^I$ satisfying the algebra\footnote{  Reinstating $\ka$ would lead to $$[\bfJ^I,\bfJ^J] = \ka \,\epsilon^{IJ}{}_K \bfJ^K. $$ } \eqref{generalcom}
\be\label{Jcom}
[\bfJ^I,\bfJ^J] = \epsilon^{IJ}{}_K \bfJ^K. 
\ee
The dual algebra $\cs^*$ is the algebra with generators 
\be
\tau_I \equiv \bfP_I+ n^J \epsilon_{IJ K}\bfJ^K = \bfP_I + (n\times \bfJ)_I 
\ee
which satisfy the $\an$ algebra commutation relations
\be\label{Pcom}
[\tau_I,\tau_J]= C_{IJ}{}^K \tau_K   \quad \textrm{with } \quad
C_{IJ}{}^K = \sigma(n_I\delta_J^K - n_J\delta_I^K) .  
\ee
With our specific choice $n^I=(0,0,s\sigma \sqrt{|\Lambda|})$ and $\eta_{IJ}=\mathrm{diag}(+,-s\sigma,-s)$, we get an algebra which is independent of $\sigma$ and $s$:
the Lie algebra $\an$ given by 
\be
[\tau_1,\tau_2]=0,\qquad [\tau_3,\tau_1]= \sqrt{|\Lambda|} \tau_1,\qquad
[\tau_3,\tau_2]= \sqrt{|\Lambda|} \tau_2.
\ee

 The symmetric pairing is simply
\be
 \la \tau_J, \bfJ^I \ra=\delta^I_J=\la \bfJ^I,\tau_J\ra , \qquad 
\la \bfJ^I, \bfJ^J\ra =0=\la \tau_I,\tau_J\ra.
\ee
We emphasize that the structure constant $C_{IJK}$ is not cyclic as $\epsilon^{IJK}$. 
The last structure constant is the mixed one 
\be \label{crosscom}
[\bfJ^I, \tau_J] = C_{JK}{}^I \bfJ^K +\epsilon^{I}{}_{J}{}^K \tau_K,
\ee
which is uniquely determined from (\ref{Jcom},\ref{Pcom}) by the Killing form defining property 
$
\la [X,Y],Z\ra=\la X,[Y,Z]\ra.
$

\medskip 

The Drinfeld double decomposition of $\ds_{\sigma s}$  is given by the Iwasawa decomposition
\be
\ds_{\sigma s}=\su_{\sigma s} \bowtie \an\sim \an \bowtie \su_{\sigma s}. 
\ee
Such an Iwasawa decomposition does  not exist for $\ds_{++}\sim\so(4)$, which is why we do not consider it.  The cross commutator \eqref{crosscom}  includes an action 
of $\su_{\sigma s}$  on $\an$
of and retro-action of $\an$ on $\su_{\sigma s}$. We can isolate the different actions,   by considering the projection of the cross commutator \cite{Majid:1996kd}.
 \bea\label{not0}
 \bfJ^I\rhd  \tau_J \equiv [\bfJ^I, \tau_J]_{{\an}}&= \epsilon^{I}{}_{J}{}^K \tau_K, \quad& \bfJ^I\lhd  \tau_J \equiv [\bfJ^I, \tau_J]_{{\su}}= C_{JK}{}^I \bfJ^K \\
  \tau_J \lhd  \bfJ^I  \equiv [ \tau_J, \bfJ^I]_{{\an}}&=-  \bfJ^I\rhd  \tau_J , \quad &\tau_J \rhd \bfJ^I  \equiv [ \tau_J, \bfJ^I]_{{\su}}= -\bfJ^I\lhd  \tau_J . \label{not1}\eea
 The relations \eqref{Jcom}, \eqref{Pcom} and \eqref{crosscom} are the counterparts of \eqref{algebra}. They are the defining relations of $\ds_{\sigma s}$ as a Drinfeld double of $\su$ (with a non-trivial cocycle) \cite{Chari:1994pz}. 
Again, we emphasize that with the convention we took, the $\an$ sector always singles out the direction 3 and is independent of $(\sigma,s)$. As we will see later, the function algebra over the Lie group $\AN$ is isomorphic to the enveloping algebra $\UQ$ which is always defined with the preferred direction 3.  
\medskip 
Given $\alpha,\beta \in \su$ and $\phi,\psi \in \an$ we can summarize the Drinfeld double algebra as \cite{Majid:1996kd}
\bes \label{sub lie algebra}
&&[\alpha,\beta ] = (\alpha\times \beta)_I \bfJ^I,\qquad
[\phi,\psi]= ((\phi\times \psi)\times n)^I \tau_I \\
\label{cross terms} &&
\alpha \rhd  \phi = (\alpha \times \phi)^I\tau_I= -\phi\lhd\alpha, \quad \alpha\lhd \phi =  (\phi\times( \alpha \times n))_I \bfJ^I=-\phi\rhd\alpha.
\ees
and the cross-commutator is
\be
[\alpha ,\phi]=  \alpha \rhd  \phi+ \alpha\lhd \phi = -  \phi  \lhd \alpha - \phi\rhd  \alpha ,
\ee 
 in accordance with \eqref{not0}, \eqref{not1}  and \eqref{newcom2}.
\medskip 

\paragraph{Role of the matrix $\mr$.  } As we have seen, the source of the deformation of the boundary symmetry algebra 
is contained\footnote{ This should not be confused with the  r-matrix 
$r$ of the double introduced later.} in the ``little" r-matrix $\mr_{IJ} =\epsilon_{IJK}n^K$
that sources the canonical transformation.

Let us clarify  the algebraic role of $\mr$. This r-matrix  can be seen as  building up the $\an$ Lie algebra structure from the $\su$ Lie algebra. First let us define the  two operators 
$\mr_\pm : \an \to \su$, given by 
\be
\mr_\pm( \tau_I) =\mr_{IJ} J^J \pm \sqrt{\sigma \Lambda} \eta_{IJ} J^J,
\ee
we can recover the $\an$ Lie bracket from the $\su$ bracket. 
\be
[\phi,\psi]_\an= [\mr_+(\phi),\mr_+(\psi)]_\su-[\mr_-(\phi),\mr_-(\psi)]_\su.
\ee
Moreover these operators are \emph{Lie algebra morphisms}. 
Given two elements $\phi,\psi \in \an$ we have  
\be
 \mr_\pm([\phi,\psi]_\an)=  [\mr_\pm(\phi) , \mr_\pm(\psi)]_\su.
\ee
This morphism property  is equivalent to the identities $-n^2=\Lambda $ and 
\bea
(\phi\times n) \times (\psi \times n) - \sigma n^2 (\phi\times \psi) 
&=& ((\phi\times \psi) \times n)\times n \cr
\phi\times (\psi \times n) - \psi\times (\phi \times n)
&=& (\phi\times \psi)\times n.
\eea
which are consequences of the cross-product 
equality
$
(\alpha\times \beta)\times \gamma = \sigma[ (\alpha\cdot \gamma)\beta -\alpha (\gamma\cdot \beta) ].$
This key property of the matrix $\mr$ goes back to the work of Semenov-Thian-Shansky
\cite{SemenovTianShansky:1985my,SemenovTianShansky:1993ws}.

\medskip 

\paragraph{Gauge theory for a Drinfeld double algebra.  }
The frame field is now  valued in $\an$, $e\equiv e^I\tau_I$, whereas the connection $\tom$  has still value in $\su$, $\tom\equiv \tom^I \bfJ_I$. We can rewrite the momentum and angular momentum densities,
repectively $\tPp=\tPp^I\bfJ_I\in \su $ and $ \tSp=\tSp^I\tau_I\in \an$ as objects valued in the different subalgebras   and in terms of their respective Lie brackets and actions. 
Hence we can rewrite the equations of motion \eqref{motioneom} as 
{\fin \bes
\tPp&=& \rd\tom +\demi [ \tom ,  \tom ] +   \tom \lhd e,\label{gen curv} \\
 \tSp&=&\rd e
  +\demi [e , e] + \tom {\rhd} e . \label{gen tor}
\ees}
We can also rewrite the covariant derivatives \eqref{transfosu2} and \eqref{transfoan1} in the different directions. For some scalar fields, $\alpha=\alpha^I\bfJ_I \in \su$ and $\phi=\phi^I\tau_I\in \an$, we have  
 {\fin  \bes
 D \alpha &=&  \rd \alpha  + [\tom,\alpha  ] + e\rhd  \alpha,   \nn \\ 
 &=&\rd \alpha  + \tom \times \alpha  + e \times( n\times \alpha)\\
 \tilde{D} \phi &=& \rd \phi+[ e ,\phi ] +  \omega  \rhd \phi      \nn\\
  &=&  \rd \phi+ (e\times \phi)\times n + \omega \times \phi.
 \ees
  Another way to recover these relations is to consider the total connection $\cA = \tom + e$ with value in $\ds$, as we do in Appendix \ref{big-guy}.}
 
 One can check that the covariant derivatives satisfy the metric compatibility condition 
\be
\rd (\alpha \cdot \phi) =  D\alpha\cdot\phi +   \alpha \cdot\tilde{D}\phi. 
\ee
 As anticipated in \eqref{transfosu2} and \eqref{transfoan2}, the symmetry transformations, parametrized by either $\alpha\in \su$ or $\phi\in \an$, can be specified in terms of these new covariant derivatives. 
  \be
\begin{array}{lll}
 \delta'_{\alpha} \tom = D \alpha  , &\,\,&  
\delta'_{\alpha} e = e \lhd \alpha,
  \label{transf1}  \\
 \delta'_\phi \tom =    \tom \lhd \phi , 
 &\,\,&  \delta'_\phi e = \tilde{D} \phi.
 \end{array}
 \ee
These imply the following transformations for the momentum densities,
\be\label{transf F}
\begin{array}{lll}
\delta_{\alpha} \tPp =  [\tPp, \alpha]  + \tSp\rhd \alpha    , &\qquad& \delta_{\alpha} \tSp= \tSp \lhd \alpha   \\
 \delta_\phi \tPp=     \tPp\lhd  \phi  , & \qquad& \delta_\phi \tSp=    [\tSp, \phi] + \tPp \rhd \phi .
 \end{array}
\ee
It is worth noticing that these transformations now have a symmetric expression, since we have an action of $\su$ on $\an$ and a retro-action
of $\an$ on $\su$. 

\medskip
 


Finally, we use the Killing form and the fields with value in their respective algebra to define the symplectic form that we are going to discretize in the next section. 
\be
\Omega =\int_\Sigma \la \delta e \curlywedge \delta A \ra = \int_\Sigma \la \delta e \curlywedge \delta \tom \ra.
\ee

\section{Recovering the deformed loop gravity phase space}\label{sec:main}
We intend to use now the recent understanding behind the notion of discretization of gauge theories \cite{Donnelly:2016auv}.  Such discretization consists in two steps,  a \textit{subdivision} and then by a \textit{truncation} of the degrees of freedom. We will use this to derive the discretized symplectic form, which will allow us to identify the discretized phase space variables. The quantization of such variables will make obvious how the quantum group structure appears.  

\subsection{Subdivision and truncation}

\begin{figure}[h]
\begin{center}
\includegraphics[scale=.5]{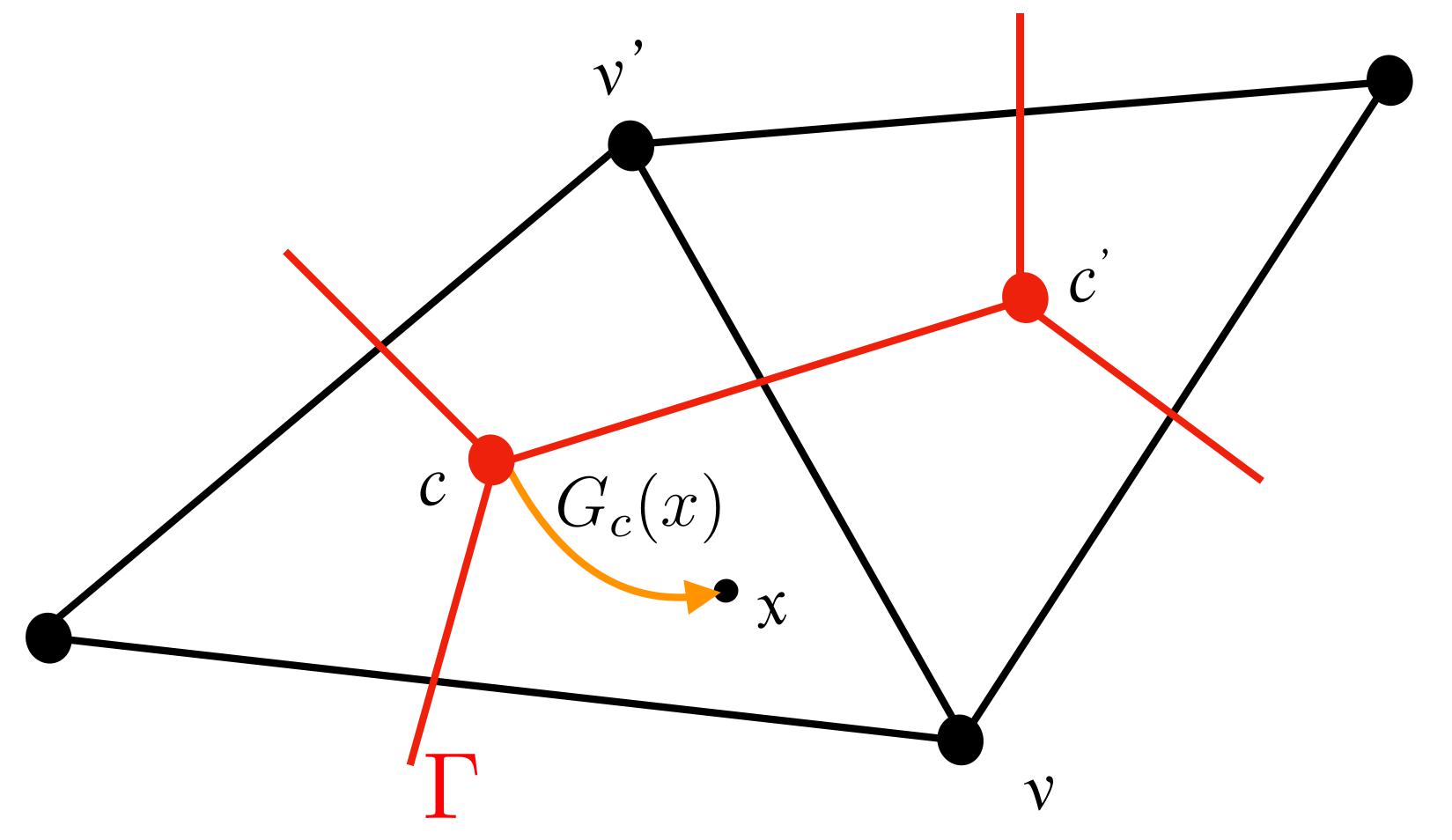}
\caption{The two subregions/triangles $c^*$ and $c'^*$ with their respective reference point/center $c$ and $c'$.  $\Gamma$ is the dual 2-complex. The segment $[cc']$ forms a link, dual to the edge $[vv']$ shared by $c^*$ and $c'^*$.  The $\AN$ and $\SU$ holonomies $\ell_{cx}$ and $h_{cx}$ are based at $c$ and go to a point $x$ in the cell $c^*$. These holonomies can be put together as a single $\DS_{\sigma s}$ holonomy $G_c(x)=\ell_{cx}h_{cx}$.}
\label{loop04}
\end{center}
\end{figure}

By subdivision, we mean  that we  decompose the (2d) Cauchy data slice $\Sigma$ into a collection of subregions. This provides a \emph{cellular decomposition } of space in terms of cells of different dimensions. The cells of maximal dimensions are denoted $c_i^*$, where $i$ labels the  cell which is dual to the center $c_i$, see Fig. \ref{loop04}. In terms of this subdivision, the symplectic form becomes 
\be
\Omega = \int_\Sigma \la \delta e \curlywedge \delta \tom \ra = \sum_i\int_{c_i^*} \la \delta e \curlywedge \delta \tom \ra.
\ee
To proceed to the evaluation of $\Omega$, we are going to perform a \textit{truncation} of the degrees of freedom, which is in a way the core of the discretization process. We will assume that any matter degrees of freedom are localized on the vertices $v$ of the triangulation.  A proper treatment of such defects could be done as in \cite{Freidel:2018pbr, Shoshany:2019ymo}. However here we will neglect them and leave for later their careful study.

Truncation refers to the fact that then in each subregion one choses a particular vacuum state or a particular family of solution of the constraints.
{\fin \bes 
0&=& \rd\tom +\demi [ \tom ,  \tom ] +    \tom \lhd e, \label{defCurv} \\
0&=&\rd e +\demi [e , e] + \tom {\rhd} e \label{sols}.\label{defGauss}
\ees}
Once this is done, the systems attached to subregions carry representations of the boundary symmetry group. The choice of discretisation scheme is achieved once we choose a representation of the boundary symmetry.

Let us identify the solutions of \eqref{sols} in a subregion  $c^*$. For this, it is convenient to consider a $\ds_{\sigma s}=\su_{\sigma s}\bowtie \an$ valued connection $\cA = \tom + e$. The associated curvature tensor is given by (see Appendix \ref{big-guy})
{\fin \be\label{r3}
\cF = ( \rd\tom +\demi [ \tom ,  \tom ] +   \tom \lhd e)_I\bfJ^I +  (\rd e +\demi [e , e] + \tom {\rhd} e )_J\tau^J.
\ee}
The gauge transformations for the connection $\cA$ are given in terms of the group $\DS_{\sigma s}\cong\SU_{\sigma s}\bowtie\AN$. This splitting is in general only local, except for the cases $\DS_{+,-}=\SL(2,\C)\cong\SU(2)\bowtie\AN$ (Euclidean case with $\Lambda<0$)  and when $\Lambda=0$, where the splitting is global. For simplicity, we only focus on the connected component to the identity. 

Demanding that \eqref{defCurv} and \eqref{sols}  are satisfied is the same as demanding that the connection $\cA$ is flat, hence it has to be pure gauge. Let us consider the $\DS_{\sigma s}$ holonomy $G_c(x)$ connecting a reference point $c$ in $c^*$ to a point $x$ still in $c^*$ (see Fig. \ref{loop04}).  
In the  connected component to the identity, we have a unique decomposition of $G_c(x)$ as 
$G_c(x)=\ell_{c}(x) h_{c}(x)$ where $\ell_{c}(x)\in \AN$ and $h_{c}(x)\in \SU$.  

We will often omit the $x$ dependence in the notation. The solutions to the constraints are given by
\be \cA_{|_{c^*}} = \tom_{|_{c^*}}+e_{|_{c^*}}= G_{c}\mone \rd G_{c} = (\ell_{c}h_{c})\mone \rd (\ell_{c}h_{c}),\ee
which in terms of components give (we recall that the Lie algebra $\an$ is not stable under the adjoint action of $\SU$),
\bes 
\tom_{|_{c^*}} & =&  h_{c}^{-1} \text{d}h_{c} \: + \: \left(h_{c}^{-1} (\ell_c^{-1} \text{d}\ell_{c}) h_{c}\right)_{|_{\su}}\label{con1}\\
  e_{|_{c^*}}  &=&  \left(h_{c}^{-1} (\ell_{c}^{-1} \text{d} \ell_{c}) h_{c}\right)_{|_{\an}} \,. \label{con2}
\ees
When considering an infinitesimal transformation, we recover the transformations \eqref{transf1} for $e=0$ and $\tom=0$.  Also these solutions are the deformed version of the standard discrete picture with $\Lambda=0$ (and $n=0$), 
\begin{equation}
\omega_{|_{c^*}}  = h_{c}^{-1} \text{d}h_{c}   , \quad e_{|_{c^*}}  = h_{c}^{-1}\rd X \, h_{c}, \quad \textrm{with } \ell \equiv X\in \R^3.
\label{standard 1}
\end{equation}

Before identifying the truncated symplectic form, it will be convenient to rewrite the restriction $\Omega_{|_c}$ of $\Omega$ to the cell $c^*$ as
\be
\Omega_{|_c}= \int_{c^*} \la \delta e \curlywedge \delta \tom \ra = \tfrac12  \int_{c^*} \la   \delta \cA \curlywedge \delta \cA \ra. 
\ee
The truncation then imposes  that $\cA=G\mone_c\rd G_c $. 
\be
\Omega_{|_c}\approx \Omega_c\equiv  \demi \int_{c^*}\la  \delta (G_c^{-1} \rd G_c) \curlywedge \delta (G_c^{-1} \rd G_c) \ra = \delta \Theta_c,  
\ee 
where $\approx$ means we went on-shell, ie we truncated the number of degrees of freedom. $\Theta_c =\demi \int_{c^*} \la G_c^{-1} \rd G_c \curlywedge \delta (G_c^{-1} \rd G_c) \ra$ is the truncated symplectic potential. 

\medskip

The next steps will consist in evaluating $\sum_i \Omega_{c_i}= \sum_i \delta \Theta_{c_i} $ in order to identify the discretized variables and their phase space structure.

\medskip

An important first step is to realize that what is relevant is actually the boundary data of the subregion (as we could guess already from the charge analysis in the continuum).  We will be using   extensively from now on the notation 
\be\label{defD}
\Dr u :=\delta uu \mone,\qquad \Dl u :=u \mone \delta u
\ee  for some group element $u$. 
$\Dr u$ is right invariant $\Dr(ug)=\Dr u$ and $\Dl u $ is left invariant 
$\Dl(gu) =\Dl u $, for a field independent group element $\delta g=0$.

\begin{proposition} \label{prop1}In the component connected to the identity, where  $\cD_{\sigma s}=\SU \bowtie \AN\ni G =\ell_c h_c$,  there exist a boundary symplectic potential $\vartheta$ and a boundary Lagrangian $L_\pa$ given by
\be 
\vartheta:= -  \la \ell_c^{-1} \rd \ell_c , \Delta h_c \ra ,
\qquad
 L_\pa:= \tfrac12  \la \rd h_c h_c^{-1} {\wedge}   \ell_c^{-1} \rd \ell_c\ra,
 \ee 
such that  $\Theta_c$ decomposes as a sum of a total derivative and a total variation 
\be
\Theta_c = \int_{c^*}  \left(\rd \vartheta + \delta L_\pa \right).
\ee
As a corollary we have that $\Omega_c = \delta \Theta_c = \int_{c^*} \rd \delta \vartheta =  \int_{\partial c^*}  \delta \vartheta $ is a pure boundary term.

\noindent $\blacksquare$
\end{proposition}

Let us prove this proposition. We will omit the index $c$ to simplify the notation. Some useful relations are given by 
\bea
G^{-1} \rd G &=& h^{-1} \rd h + h^{-1} (\ell^{-1} \rd \ell) h  \\
\delta (G^{-1} \rd G) &=&   h^{-1}  \rd \Delta h h+\delta ( h^{-1} (\ell^{-1} \rd \ell)h) \cr
&=&
h^{-1} (\rd \Delta h + [(\ell^{-1} \rd \ell),\Delta h ]+ \delta (\ell^{-1} \rd \ell)  ) h 
\eea
Using these, we directly get 
\bea
2\Theta &=& \int \la G^{-1} \rd G \wedge\delta (G^{-1} \rd G) \ra\cr
&=&  \int \la (\ell^{-1} \rd \ell)  \wedge (\rd \Delta h + [(\ell^{-1} \rd \ell),\Delta h ]) \ra
+ \la h^{-1} \rd h , \delta ( h^{-1} (\ell^{-1} \rd \ell)h)\ra 
 \cr
&=&\int  - \rd \la (\ell^{-1} \rd \ell)  \wedge  \Delta h \ra + \frac12 \la [\ell^{-1} \rd \ell, \ell^{-1} \rd \ell]  \wedge  \Delta h \ra \cr
&& + \int \delta  \la \rd h h^{-1},  \ell^{-1} \rd \ell\ra - 
\la  \rd \Dr h,  (\ell^{-1} \rd \ell)\ra
\cr&=&\int - 2\rd \la (\ell^{-1} \rd \ell)  \wedge  \Delta h \ra 
+ \delta  \la \rd h h^{-1}\wedge  (\ell^{-1} \rd \ell)\ra.
\eea
which establishes the result.

\medskip
 
%
%
%
Therefore the symplectic form associated with a cell $c^*$ can be written as
a sum of boundary edge contributions
\be
\Omega_{|_c^*}=\delta \Theta_{|_c^*}=\delta \int_{c^*}   \la  e \wedge \delta \tom\ra \approx \Omega_c  
= \sum_{e \in \pa c^*} \Omega_{c}^{e},
\ee
where each contribution in the sum is given by 
\be \label{Thetac}
\Omega_c^e =\delta \Theta_c^e,\quad \Theta_c^e := -\int_e \la\ell^{-1}_c \rd \ell_c , \Dr h_c  \ra.
\ee

\subsection{From holonomy to ribbon and Heisenberg double}

\subsubsection{From holonomies to ribbons }
The different subregions $ c^*$ and $ c'^*$ share some common boundaries. This common boundary is referred to as an edge $e$. This means that the variables evaluated on the edge can be related through transformations relating the different frames associated to each  triangle. As we will see this will generate some simplifications in the total symplectic form $\sum_{i}\Omega_{c_i}$.    
 
Let us now focus  on two cells $c^*$ and $c'^*$,  sharing the edge $e=[vv']$, where $v$ and $v'$ are vertices of the cellular decomposition. 
As a set we have $c^*\cap c'^*=[vv']$, in addition $[vv']$ possesses an orientation induced by  the orientation of $c$, see Fig \ref{loop04}.  
We have two contributions, for  the edge $[vv']$ coming from the two cells sharing $[vv']$. 
\be \label{split 2}
\Omega_{c c'} \equiv  \Omega_{ c}^{[vv']} +  \Omega_{ c'}^{[v'v]}= 
\Omega_{ c}^{[vv']} -  \Omega_{ c'}^{[vv']},
\ee
where the sign changed because the edge $[vv']$ has a different orientation depending whether it is belonging to the boundary of 
$ c^*$ or $ c'^*$.
On the boundary $[vv']$, the different fields  can be  combined as $\DS_{\sigma s}$ holonomies  $G_{c_i} = \ell_{c_i} h_{c_i}$, with 
$\ell_{c_i} \in \AN$ and $h_{c_i} \in \SU_{\sigma s}$,  are related by a $\DS_{\sigma s}$-transformation. 
The continuity equation states that the connection evaluated on $[vv']$ can be expressed either from the perspective of the frame of $c^*$ or the one of $c'^*$. 
\be
\cA(x) =(G_c^{-1}\rd G_c)(x) = (G_{c'}^{-1}\rd G_{c'})(x), \qquad x\in [vv'].
\ee
This differential  equation can be integrated.  Indeed, the group elements  $ G_{c}(x)\equiv G_{cx}$ and $ G_{c'}(x)\equiv G_{c'x}$ are evaluated at the same point $x\in [vv']$ and since the connection is flat,   there exists an holonomy $\cG_{c'c}=L_{c'c} H_{c'c}$ such that $G_{c'}(x)=\cG_{c'c} G_{c}(x)$. 
 Note that for any given holonomy $G_{xy}$ connecting $x$ to $y$, we take the convention $G_{yx}\equiv G_{xy}\mone $. 

\begin{figure}[h]
\begin{center}
\includegraphics[scale=.30]{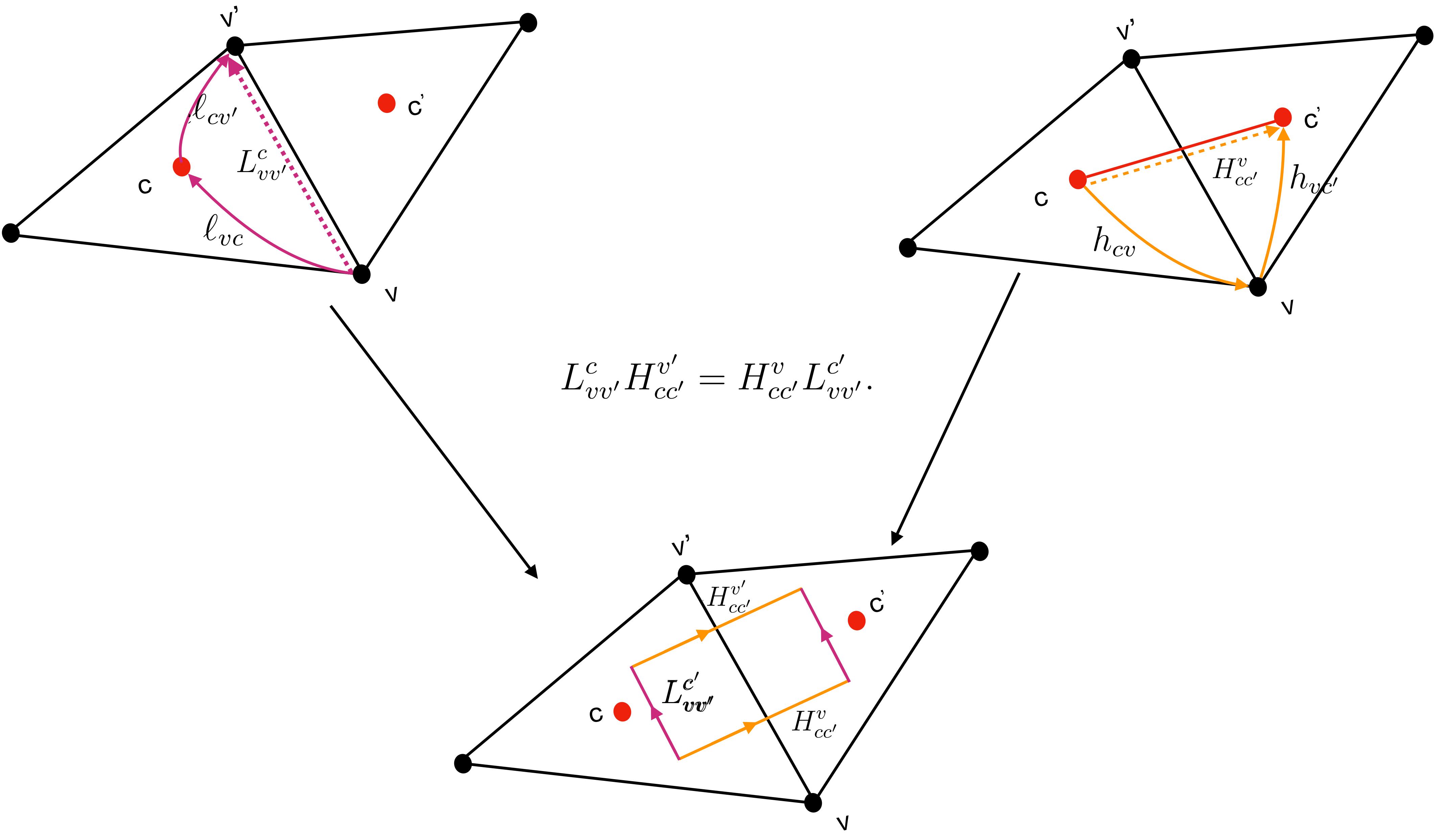}
\caption{The constraint \eqref{iwatop} provides  the natural way to define a ribbon structure associated to each link $[cc']$. It encodes that the holonomy around the ribbon is trivial.   }
\label{loop03}
\end{center}
\end{figure}

The differential continuity equation is 
\be
\pa_x G_{c x}G_{x c'}=0.
\ee
for $x\in [vv']$. This implies the integrated continuity condition
\be\label{intcont}
G_{c v}G_{v c'}=  G_{c v'}G_{v' c'}.
\ee
Using the left Iwasawa decomposition $G_{cx}=\ell_{cx} h_{cx}$  in the cell $c^*$ and the right one\footnote{{\fin We note that since the inverse is an antihomomorphism, $G_{yx}=\ell_{yx} h_{yx} \dr G_{xy}\mone=h_{xy}\mone \ell_{xy}\mone= G_{yx} = h_{yx}\ell_{yx}$, the right decomposition of the inverse is the analogue of the left decomposition. 
}} $G_{xc'}=h_{xc'} \ell_{xc'}$ in $c'^*$, we can rewrite this condition  as 
\be
 \ell_{cv} h_{cv} h_{vc'} \ell_{vc'}=  \ell_{cv'} h_{cv'} h_{v'c'} \ell_{v'c'}
\quad \Leftrightarrow \quad  h_{cv} h_{vc'} \ell_{vc'}\ell_{c'v'}=  \ell_{vc}\ell_{cv'} h_{cv'} h_{v'c'}.
\ee
In other words once we introduce the {\it triangular holonomies}
\be
 L_{vv'}^c\equiv \ell_{vc}\ell_{cv'}\in \AN, \qquad H_{cc'}^v \equiv h_{cv} h_{vc'}\in \SU_{\sigma s},
\ee
and we can express the integrated continuity equation (\ref{intcont}) as the  
\emph{ribbon structure}, see Fig. \ref{loop03},
\be \label{iwatop} 
  L_{vv'}^c H_{cc'}^{v'} = H_{cc'}^v L_{vv'}^{c'}.
\ee
The triangular holonomies are the classical analogues of Kitaev's triangle operators \cite{Kitaev1997, triangle}. 

\subsubsection{Heisenberg double/phase space associated to a link} 

Having such a ribbon structure points for a natural symplectic form \cite{Alekseev_1994}. In fact we are going to prove that the explicit evaluation of $\Omega_{c c'}$, defined in \eqref{split 2}, is the natural symplectic form making $\DS_{\sigma s}$ a Heisenberg double, the generalization of the notion of cotangent bundle as a phase space \cite{Alekseev_1994}.

\medskip

\begin{theorem}\label{thm}
The  symplectic form associated to a link $[cc']$ is given by 
\bes\label{main}
\Omega_{c c'}&=& \Omega_{c}^{{[vv']}} -  
\Omega_{ c'}^{[vv']} = \demi \left(\la \Dr H^v _{cc'}\, \wedge\,  \Dr L^{c}_{vv'}\ra + \la\Dl H^{v'}_{cc'}\,\wedge\,    \Dl L^{c'}_{vv'}\ra \right).
 \ees
 $\blacksquare$
 \end{theorem}

 \medskip
The  proof of this result is presented in section \ref{proof}.  This theorem can be seen as the main result of the paper. 
Before proving the theorem it can be instructive to check that $\Omega_{c c'}$ is indeed closed  \cite{Alekseev_1994}. 

 For notational simplicity, let us omit the indices  and lets assume that 
 $\ell, \tl\in\AN$, $h,\th\in\SU$ are such that  they form a ribbon structure
\be { G\equiv \ell\, h= \th\,\,\tl}. \label{iwa0}
\ee
The  2-form    $\Omega_{c c'}=\Omega$ can then be written as  
\be\label{goal}
\Omega= \frac12 \Omega_L +\frac12 \Omega_R,\qquad
\Omega_L :=  \la  \Dr \th \wedge \Dr \ell  \ra
,\qquad
\Omega_R :=  \la  \Dl h \wedge  \Dl \tl \ra  ,
\ee

The variation of this equation implies that $ \Delta G= \Dr \ell + G  \Dl h G^{-1} $, also that 
$ \Delta G= \Dr \th + G  \Dl \tl G^{-1}$ and the identity
\bea
\Dr \ell- \Dr\th    &=&  G (\Dl \tl -  \Dl h )G^{-1}.
\eea
Since $\delta \Dr \th = \Dr \th\wedge \Dr \th$, and  
$\delta \Dl h=- \Dl h\wedge \Dl h$, one finds that
\bea
\delta \Omega_L &=&\la  (\Dr \th-\Dr \ell) \stackrel{\wedge}{,} \Dr \th\wedge \Dr \ell \ra
\cr
&=& \frac13  \la  (\Dr \th-\Dr \ell) \stackrel{\wedge}{,} (\Dr \th-\Dr \ell)\wedge (\Dr \th-\Dr \ell) \ra
\eea
We used in the second equality the fact that $\AN$ and $\SU$ are isotropic.
We  find a similar result for $\Omega_R$ with $(\Dr \th-\Dr \ell)$ replaced by 
$-(\Dl \tl -  \Dl h )$ and therefore $\delta \Omega_R =-\delta \Omega_L$, and $\Omega$ is closed.
Hence the  Poisson bracket  associated to $\Omega_{c c'}$ satisfies the Jacobi identity. 
%
This phase space structure generalizes the usual notion of cotangent bundle.

%

\subsection{Drinfeld double as symmetry of the Heisenberg double}\label{sec:sym}

\paragraph{Match pair of groups.}  We recall that the decompositions of  $\DS_{\sigma s}$ into $\AN$ and $\SU_{\sigma s}$ provide the definitions of actions of $\AN$ on $\SU_{\sigma s}$  and vice versa. This allows to see $\DS_{\sigma s}$ as a matched pair of groups \cite{Majid:1996kd}.
\bes
\ell h = \th\tl = (\ell \rhd h )(\ell\lhd h ) \Rightarrow \ell \rhd h \equiv \th, \quad \ell\lhd h \equiv \tell\\
\th\tl = \ell h = (\th\rhd \tl ) (\th\lhd \tl )\Rightarrow \th\rhd \tl \equiv \ell, \quad \th\lhd \tl \equiv h.
\ees
Some of the compatibility properties of the actions are as follows.
\bes
&&1\lhd h =1, \,\, \ell \lhd(h_1h_2)= (\ell \lhd h_1)\lhd h_2, \,\, (\ell_1\ell_2)\lhd h = (\ell_1\lhd(\ell_2\rhd h))(\ell_2\lhd h)\nn\\
&&\ell\rhd 1 =1, \,\, \ell \rhd(h_1h_2)= (\ell \rhd h_1)((\ell\lhd h_1)\rhd h_2), \,\, (\ell_1\ell_2)\rhd h = \ell_1\rhd(\ell_2\rhd h)\nn\\
&& (h\mone \rhd\ell\mone) = \tl\mone =( \ell \lhd h )\mone, \quad (h\mone \lhd\ell\mone) = \th\mone =( \ell \rhd h )\mone \label{crossaction4}
\ees
where we used in the last line the inverse of \eqref{iwa0}, namely $h\mone \ell\mone = \tl\mone \th\mone$. We have similar properties for the other actions in terms of $\th$ and $\tl$. 

\paragraph{General action of $\DS_{\sigma s}$ on itself.}  
The  Heisenberg double is defined in terms of the group $\DS_{\sigma s}$. The group $\DS_{\sigma s}$ acts on the left  (or on the right) on itself. 
\bes\label{symactions}
\begin{array}{ccc}\DS_{\sigma s}\times \DS_{\sigma s} &\dr& \DS_{\sigma s}\nn\\
(G' ,G)&\dr& G' G \end{array} \qquad \begin{array}{ccc}\DS_{\sigma s}\times \DS_{\sigma s} &\dr& \DS_{\sigma s}\nn\\
(G',G)&\dr& G G' \end{array}.  
\ees
Using either of the left or right decompositions $G=\ell h = \th \tell$,  and the left decomposition for $G'=\ell' h'$, $\ell'\in\AN, \, h'\in \SU_{\sigma s}$, we have, for the left action, 
\bes
G'G=  \ell' h' \, \ell h= [\ell' (h'\rhd \ell')] [(h' \lhd \ell)h]= \ell' h'   \th \tell = (\ell' \rhd (h' \th))
(\ell' \lhd (h' \th)) \tell. 
\ees
The left and right actions of $\DS_{\sigma s}$ on itself  encode the natural phase space symmetry actions and provide a discretization of the symmetries generated by the charges \eqref{chargedef}.

\medskip 

\paragraph{Rotations on the left.} Let us consider the infinitesimal transformations associated to left transformations (the right transformations are obtained in an analogous manner). 

Let us first look at the infinitesimal (left) action $\delta^L_\alpha$ of the rotations $h'\sim 1 + \alpha$, $\alpha\in\su$ on $G\in \DS_{\sigma s}$.
\be \label{rotsymleft} 
h'\act G = h' G \sim (1+\alpha )G \textrm{ with } G=  \ell h =  \th \tell 
\ee
We deduce then the easy transformations, 
\bes\label{whererot}
\delta^L_\alpha G= \alpha G, \quad 
\delta^L_\alpha  \th   = \alpha  \th  , \quad \delta^L_\alpha \tell = 0.
\ees
The other transformations, $\delta^L_\alpha  h  ,  \delta^L_\alpha \ell $,  require a bit more work.
We have 
\bes
h' \,\ell\,  h = (h' \rhd\ell) \, (h' \lhd\ell) \,  h \rightarrow \left| \begin{array}{l} h' \rhd\ell = h' \,\ell \, (h' \lhd\ell)\mone  
= h' \,\ell\,  (\ell\mone \rhd (h')\mone) 
\\
h'\rhd h= (h' \lhd\ell) \,  h
 \end{array}\right. . 
\ees
 So at the infinitesimal level\footnote{Note that we have  
\be\label{v to alpha}
(h' \lhd\ell)\mone 
 =   (\ell\mone \rhd h'{}\mone) \Rightarrow  - (\alpha \lhd\ell) 
 =   - (\ell\mone \rhd \alpha). \nn
\ee}, we have 
\bes 
\delta^L_\alpha \ell &= &
\alpha \ell - \ell (\alpha \lhd \ell) , \label{infi rot ell}\\
\delta^L_\alpha h &=& (\alpha \lhd\ell) \,  h.\label{infi rot h}
\ees
Since we deal with  a match pair of groups, due to the action and back action we can have a twisted compatibility relation with the product  \cite{Majid:1996kd}. In particular for the action on the $\AN$ sector we have,
\bes\label{braiding}
h\rhd (\ell_1\ell_2)= (h\rhd\ell_1) ((h\lhd\ell_1)\rhd \ell_2)\dr  \delta_{\alpha} (\ell_1\ell_2)= (\delta_{\alpha} \ell_1)\ell_2 +  \ell_1 (\delta_{\alpha \lhd \ell_1}\ell_2).
\ees
The action \eqref{infi rot ell} satisfies such condition.
\bes
\delta_{\alpha} (\ell_1\ell_2)&=&  
\alpha \ell_1\ell_2 - \ell_1\ell_2 (\alpha \lhd (\ell_1\ell_2))\nn\\
&=&\{ ( \alpha \ell_1 ) - \ell_1 (\alpha \lhd\ell_1) \}\ell_2 + \ell_1\{ (\alpha \lhd\ell_1) \ell_2 - \ell_2 (( \alpha \lhd\ell_1)\lhd \ell_2)\} \nn\\
&=&(\delta_{\alpha} \ell_1)\ell_2 +  \ell_1(\delta_{\alpha \lhd\ell_1}\ell_2).
\ees

\medskip

\paragraph{Charge for the rotations on the left.} {\fin In the continuum picture we have identified the charges $J'$ generating the rotational symmetry. The following proposition determines the corresponding charge in the discrete picture. 
\begin{proposition}\label{prop:sym1}
The triangular holonomy $\ell=L^{c}_{vv'}$ generates the infinitesimal left rotations.
\bes
\delta^L_\alpha   \ip  \Omega_{c c'}
&=& \la \alpha  \, , \, \Dr \ell \ra .\label{gausscharge}
\ees
\end{proposition}
We provide the proof in Appendix \ref{proofprop:sym1}. Geometrically this (infinitesimal) rotation is located at $c$ as it can be read from \eqref{whererot}, remembering that $\tell = L^{c'}_{vv'}$ and $\th= H^v_{cc'}$.

\medskip

\paragraph{Generating left rotations  with Poisson brackets.} 
The Poisson bracket associated to the symplectic form can be obtained by inverting the symplectic form \cite{Alekseev_1994}. We can also directly infer it  from the infinitesimal transformations. Indeed, as discussed in \cite{Babelon_1992}, since  $\ell$ is the charge of the left rotation $\delta^L_\alpha$ we can recover from the action of $\delta^L_\alpha$ on $(\ell,\tl,h,\th)$
the Poisson bracket of $\ell$ with all the other components,  using the correspondence
\be\label{babelon1}
\delta^L_\alpha \cdot = -\la \alpha\,,\,\{\ell_{1}\,,\, \cdot     \} \ell_{1}\mone \ra_{1}  , 
\ee
where we are  using here the notation $\ell_{1}:= \ell\ot 1$,  $\ell_{2}:= 1\ot \ell$ and  $\la,\ra_{1}$ means we are contracting the first sector of the tensor product.  

\begin{proposition}\label{prop:sym2}
The Poisson brackets implementing the infinitesimal transformation \eqref{babelon1}  can be conveniently written in terms of the  $r$-matrix  \cite{Semenov1992}
\be 
\Rm\equiv  - \tau_I \ot \bfJ^I.
\ee
%
and are  given by 
\bea
\poi{\ell_{1},\ell_{2}}= \com{\Rm, \ell_{1}\ell_{2}}, &\, &
\poi{\ell_{1},h_{2}}=\ell_{1} \Rm h_{2}, \\ 
\poi{\ell_{1}, \tell_{2}}= 0, &\, &
 \,\, 
\poi{\ell_{1}, \th_{2}}= \Rm \, \ell_{1} \th_{2}. \nn
 \eea
 \end{proposition}
We provide  the proof in Appendix \ref{proofprop:sym2}.}
\medskip


\paragraph{Translations on the left.} 

A similar calculation can be performed for the infinitesimal (left) translations $\delta^L_\phi$, 
$\AN\ni \ell'\sim 1 + \phi$, $\phi\in\an$.
\be \label{transsymleft} 
\ell'\act G = \ell' G \sim (1+\phi )G \textrm{ with } G=  \ell h  =  \th\tell 
\ee
We deduce again the easy transformations, 
\bes\label{wheretrans}
\delta^L_\phi G= \phi G, \quad 
\delta^L_\phi  \ell   = \phi  \ell    , \quad \delta^L_\phi h = 0 , 
\ees
and the other transformations, $\delta^L_\phi  \th  ,  \delta^L_\phi \tell $,  require a bit more work.
We have 
\bes
 \,\th \,\tell\,  = (\ell'  \rhd\th) \, (\ell'  \lhd\th) \,  \tell \rightarrow \left| \begin{array}{l} \ell'  \rhd\th = \ell'  \, \th \,  (\ell'  \lhd\th)\mone  = \ell'  \, \th  \,(\th\mone \rhd \ell'{}\mone) \\
\ell' \rhd \tell= (\ell'  \lhd\th) \,  \tell
 \end{array}\right. . 
\ees
We note that the formulae are actually very similar to the left rotations we first determined. It is natural since the construction is by essence symmetric between the $\su$ and $\an$ sectors. 

At the infinitesimal level, we have 
\bes 
\delta^L_\phi \th  &= &\phi \th - \th (\th\mone \rhd \phi) = \phi \th - \th ( \phi \lhd \th ) , \label{infi trans h}\\
\delta^L_\phi \tell  &=& (\phi \lhd\th) \,  \tell.\label{infi trans ell}
\ees
It is clear  that the action \eqref{infi trans h} satisfies a twisted compatibility condition with the product of $\SU$, since the formula is very similar to \eqref{infi rot ell}. 

\medskip



\paragraph{Charge for the translations on the left.}   {\fin In the continuum picture we have identified the charges generating the translation symmetry $P'$. The following proposition determines the corresponding charge in the discrete picture.  As one could expect, the charge generating the left translation is now given by the $\SU$ holonomy.

\begin{proposition}\label{prop3}
The triangular holonomy $\th=H^{v}_{cc'}$ generates the infinitesimal left translations.
\bes
\delta^L_\phi   \ip  \Omega_{c c'}
&=&
 -  \la \phi ,\, \Dr \th\,  \ra \label{transcharge}.
\ees
\end{proposition}
The proof of this proposition is very close to the one of Proposition \ref{prop:sym1}, thanks to the symmetric treatment between the variables $\ell \leftrightarrow \th$, $\tell \leftrightarrow h$, and sectors $\an \leftrightarrow \su$.  Geometrically this (infinitesimal) translation is based at $v$, as it can be read from \eqref{wheretrans}, remembering that $\ell= L^c_{vv'}$ and $h=H^{v'}_{cc'}$.

\medskip

\paragraph{Generating left translations  with Poisson brackets.} 
We can also derive the infinitesimal translations using the Poisson bracket. 
\be\label{babelon2}
\delta^L_\phi \cdot = \la \phi\,,\,\{\th_{1}\,,\, \cdot     \} \th_{1}\mone \ra_{1}. 
\ee
The difference of minus sign with respect to \eqref{babelon1} is due to the fact that the charges have opposite sign as one can see looking at \eqref{transcharge} and \eqref{gausscharge}. 
\begin{proposition}\label{prop:sym4}
The Poisson brackets implementing the infinitesimal transformation \eqref{babelon2}  can be conveniently written in terms of the  $R$-matrix  \cite{Semenov1992}
\be 
\Rp = J_I\ot \tau^I,
\ee
and are given by 
\bes
 \poi{\th_{1},\th_{2}}=\com{\Rp,\th_{1}\th_{2}}, \quad
\poi{\th_{1}, \tell_{2}}= \th_1\Rp \tell_2, \quad  
\poi{\th_{1}\,,\,h_{2}}=0, \quad 
\poi{\th_{1}\,,\,\ell_{2}} = \Rp \th_{1} \ell_{2}. 
\ees
 \end{proposition}
The proof is given in Appendix \ref{proofprop:sym4}.  It is very similar to the earlier proof of proposition \ref{prop:sym2} due to the symmetry between the sectors $\SU$ and $\AN$ in the different decompositions. 
}

\smallskip

A similar construction can be done for the infinitesimal right translations and rotations, which are respectively generated by $H_{cc'}^{v'}= h$ and $L^{c'}_{vv'}=\tell$ and act respectively at $v'$ and $c'$. {\fin Determining these infinitesimal transformations  allows to find the missing Poisson brackets, such as in particular
\be
\poi{h_{1},h_{2}}= -\com{\Rp, h_{1}h_{2}}, \qquad
\poi{h_{1}, \tell_{2}}= -h_1\tell_2 \Rp, \qquad \poi{\tell_{1},\tell_{2}}
= -\com{\Rm, \tell_{1}\tell_{2}} .
\ee
}
These can be obtained by the correspondence $\th^{-1} \to h$.
\smallskip
In summary we find that the Heisenberg poisson brackets when restricted to the variables
$(h,\ell)$ are\footnote{Note that since $r_+=r_-+C$, we have $-[\Rp,h_{1}h_{2}]=-[\Rm,h_{1}h_{2}] $.}  
\bea\label{Heisc}
\poi{\ell_{1},\ell_{2}}= \com{\Rm, \ell_{1}\ell_{2}}, \qquad
\poi{\ell_{1},h_{2}}=\ell_{1} \Rm h_{2}, \qquad
\poi{h_{1},h_{2}}= -\com{\Rm, h_{1}h_{2}}.
 \eea

\paragraph{Finite transformations. } We can also look at the  finite version of the left or right transformations. These are obtained from the group $\DS_{\sigma s}$ acting on itself as we have discussed earlier \eqref{symactions}. We can prove that they are phase space symmetries if we equip the group $\DS_{\sigma s}$ with another  Poisson structure, which this time is not invertible (it is however compatible with the group product of $\DS_{\sigma s}$). In this case,    $\DS_{\sigma s}$ as a symmetry group is called the Drinfled double. 
In order to write these  we note that  the r-matrices $(\Rp,\Rm)$ satisfy the relations
\be
2 r:= \Rp+\Rm, \qquad  \Rp-\Rm = C
\ee
where $C$ is the quadratic Casimir of $\ds$ and we have introduced   the  antisymmetric $r$-matrix $r$.
\bes
\textrm{\,\,\,\,Heisenberg double :} \{ G_1,G_2\} = [r,{G\ot G}]_+ = {r G\ot G} + { G\ot G} r , \label{ps}\\
\textrm{Drinfeld double :}  \{ G'_1,G'_2\} = [r , G'\ot G']_-= {r G'\ot G'} - { G'\ot G'} r , \label{drin}
\ees
with $G,G'\in \DS_{\sigma s}$. 
{\fin
The set of Poisson brackets we just derived  in \eqref{Heisc} are equivalent to the Poisson brackets \eqref{ps}. On the other hand the Poisson brackets given in \eqref{drin} are simply  \cite{Semenov1992}, 
\bea
\poi{\ell'_{1},\ell'_{2}}= \com{\Rm, \ell'_{1}\ell'_{2}}, \qquad
\poi{\ell'_{1},h'_{2}}=0= \poi{h'_{1},\ell'_{2}}, \qquad
\poi{h'_{1},h'_{2}}= \com{\Rp, h'_{1}h'_{2}}.
\eea

The left or right action of $\DS_{\sigma s}$ as a Drinfeld double on $\DS_{\sigma s}$ as a Heisenberg double  is a Poisson map \cite{Semenov1992}.  This means in physical terms that our phase space structure is covariant under the action of the Drinfeld double, which encodes some symmetry transformations equipped with a (in general non-trivial) Poisson structure. Upon quantization, the non-trivial Poisson structure becomes the relevant non-commutative/quantum group structure. Our quantum mechanical states being built from  representations of these symmetries will then  be naturally defined in terms of quantum group representations. We will come back to this point in Section \ref{qgp}.}

 \subsection{Proof of the main result} \label{proof}
 Let us prove here the main result of the paper given by theorem \ref{thm}. 
We start from the discretized symplectic form on the boundary of the cell $c$. Within any cell $c^*$ we have from Proposition \ref{prop1} that 
\bes
\Theta_{|_c}=\int_{c^*}\la e \wedge  \delta\tom \ra  \approx \Theta_{c}= \sum_{[vv']\in \pa c^*}\Theta_{c}^{[vv']},\quad
\Theta_{c}^{[vv']} =  -\int_{[vv']}\delta \la \ell_{c}\mone \rd \ell_{c} \wedge \Dr h_{c}\ra. 
\ees 
Given two cells $c^*,c'^*$ one defines 
the holonomy $G_{cc'}= L_{cc'} H_{cc'}$ and denote 
$G_{cx}= \ell_{cx} h_{cx}$, $G_{xc}= G_{cx}^{-1}$.
We also denote, for any holonomy $u_{ab}$ from to $a$ to $b$, $u\mone_{ab}=u_{ba}$. 
Given $x\in[vv']$, one defines 
\be \label{defH}
H^x_{c'c}\equiv h_{c'x}h_{xc},\qquad \tl_{cx}\equiv 
L_{cc'}\ell_{c'x}. 
\ee
Taking the variation  of  the first equation of \eqref{defH}, we get 
\bes\label{hshift}
\Dr h_{c'x}&=& \Delta H^x_{c'c} +H^x_{c'c} \Delta h_{cx} H^x_{cc'} = H^x_{c'c}(\Dr h_{cx}-\Dr H^x_{cc'} ) (H^x_{c'c})\mone,
\ees
where we have used that $\Dr H^{-1}   = - H^{-1} \Dr H H$.
Taking the differential of  the second relation in (\ref{defH}) gives
\be\label{lshift}
\ell_{c'x}\mone \rd \ell_{c'x}= \tl_{cx}\mone \rd \tl_{cx}.
\ee

The continuity equations across the edge $[vv']$ separating $c$ from $c'$ is equivalent to an  exchange relation:  
\be\label{contui1}
G_{c'c} G_{cx}= G_{c' x},\quad \Leftrightarrow\quad 
 H_{c'c} \ell_{cx}  
=  \tl_{cx} H^x_{c'c}.
\ee 
Taking the differential of the continuity equation 
\eqref{contui1}, we get
\bes
\label{diffcont}
 \tl_{cx}\mone \rd \tl_{cx} &=& 
\left(H^x_{c'c}  (\ell_{cx}\mone \rd \ell_{cx})H^x_{cc'}  + H^x_{c'c} \rd H^x_{cc'} \right). 
\ees
This relation, together with (\ref{hshift},\ref{lshift}) allows  us to relate the contribution of the cell $c'$ to the one of the  cell $c$. Denoting $\Theta_{cc'}= \Theta_{c}^{[vv']}-\Theta_{c'}^{[vv']}$ with $\Theta_c^e := -\int_e \la\ell^{-1}_{cx} \rd \ell_{cx}, \Dr h_{cx}  \ra$, see \eqref{Thetac}, one finds that 
\be
\Theta_{cc'} = -\int_{[vv']} \la \ell_{cx}\mone \rd \ell_{cx} {,} \left( \Dr H ^x_{cc'}  \right)\ra
=  \int_{[vv']} \la \tl_{cx}\mone \rd \tl_{cx} {,}\left( \Dr H ^x_{c'c}  \right)\ra. 
\ee
The second equality is due to the differential continuity equation \eqref{diffcont} 
and the identity $\Dl H   = H^{-1} \Dr H H=- \Dr H^{-1}$. 
The fact that there are two equivalent expressions for the symplectic potential simply follows from the exchange $c\leftrightarrow c'$. Under this exchange $\Theta_{cc'}$ is antisymmetric.
It is also clear from the continuity equation written as 
$\tl_{cx}^{-1} H_{c'c} \ell_{cx}  
=   H^x_{c'c} $ that under this exchange we have  $\tl_c\leftrightarrow \ell_c$. 


The variation of the differential continuity \eqref{diffcont} gives
\bea
H^x_{cc'} \delta( \tl_{cx}\mone \rd \tl_{cx})H^x_{c'c}  -  \delta (\ell_{cx}\mone \rd \ell_{cx})&=& 
 \left(  [ \ell_{cx}\mone \rd \ell_{cx}, \Delta H^x_{cc'}]  +  \rd \Dr H^x_{cc'} \right).
\eea
One can use this to  establish that
\bea
 \delta \la  (\ell_{cx}\mone \rd \ell_{cx})    \Dr H ^x_{cc'}  \ra
&=&
\la  \delta (\ell_{cx}\mone \rd \ell_{cx}) \curlywedge   \Dr H ^x_{cc'}  \ra +\la [\ell_{cx}\mone \rd \ell_{cx} {,} \Dr H ^x_{cc'}] \curlywedge   \Dr H ^x_{cc'}  \ra\cr
 &=& 
 \la H^x_{cc'} \delta( \tl_{cx}\mone \rd \tl_{cx})H^x_{c'c}  \curlywedge   \Dr H ^x_{cc'}  \ra
 \cr
 &=& - \la  \delta( \tl_{cx}\mone \rd \tl_{cx})  \curlywedge    \Dr H ^x_{c'c}  \ra.
\eea
where we have  denoted  $\curlywedge $ the variational wedge product.
This means that
\bea\label{omprime}
\Omega_{cc'}
&=& \int_{[vv']} \la  \delta( \tl_{cx}\mone \rd \tl_{cx})  \curlywedge   \Dr H ^x_{c'c}  \ra = - \int_{[vv']} \la\delta (\ell_{cx}\mone \rd \ell_{cx}) \curlywedge  \Delta H^x_{cc'}\ra .
\eea

From the variation of  the continuity equation  \eqref{contui1}
one gets  
 \bea\label{varcont}
\Dr H ^x_{cc'} &=& \Dl \ell_{cx} +\ell_{xc}\Dl H_{c'c} \ell_{cx}
+ \ell_{xc}H_{cc'} \Dl \tl_{xc} H_{c'c}\ell_{cx}\cr
&=&\ell_{c}\mone \left\{ H_{cc'}  (\Dr \tl_{c})  H_{c'c}   + \Dr H_{cc'} -\Dr \ell_c      \right\}\ell_{c},
 \eea
 where we have used that $\Dl H^{-1}=-\Dr H$.
 Similarly we have an equivalent variational continuity  identity obtained by exchanging $c\leftrightarrow c'$ and $\ell_c\leftrightarrow\tl_c$
 \bea\label{varcont2}
  \Dr H ^x_{c'c} 
&=&\tl_{c}\mone \left\{  H_{c'c}  (\Dr \ell_{c})  H_{cc'}   + \Dr H_{c'c} - \Dr \tl_c      \right\}\tl_{c} . 
 \eea
 Using these relations and the fact  that $\delta (\ell_c\mone \rd \ell_c)= \ell_c^{-1}( \rd \Delta \ell_c)\ell_c$ one can evaluate \eqref{omprime} 
  \bea
\Omega_{cc'} &=& -\int_{[vv']} \la  \rd \Dr \ell_{c} \curlywedge
\left(  H_{cc'}  (\Dr \tl_{c})  H_{c'c}   + \Dr H_{cc'}       \right)\ra, \label{om1}\\
&=& \int_{[vv']} \la  \rd \Dr \tl_{c} \curlywedge
\left(  H_{c'c}  (\Dr \ell_{c})  H_{cc'}   + \Dr H_{c'c}       \right)\ra. \label{om2}
\eea
Note that we repeatedly use the fact that the subalgebra $\su$ or $\an$ are isotropic with respect to our scalar product.
Quite remarkably the integrant of  $\Omega_{cc'}$ is a total differential.
This can be simply seen by taking  the sum of (\eqref{om1}) and (\eqref{om2}) which gives after integration 
\be \label{omdiscrete}
\Omega_{cc'} =
- \frac12 
\left.\left(  
  \la  \Dr \ell_{c} \curlywedge \Dr H_{cc'}       \ra  +
\la \Dr \ell_{c} \curlywedge   H_{cc'}  (\Dr \tl_{c})  H_{c'c}   \ra
+ \la    \Dr H_{c'c} \curlywedge \Dr \tl_{c}   \ra 
\right)\right|_{x=v}^{x=v'}. 
\ee
To evaluate this expression one recall the definition of the  triangular  holonomies
\be
L_{vv'}^c = \ell_{vc}\ell_{cv'},\qquad  L_{v'v}^{c'}= \tl_{v'c}\tl_{cv}.
\ee
Taking their variation gives
\be 
(\Dr \ell_{cv'}-\Dr \ell_{cv}) =\ell_{cv}\,(\Dr L_{vv'}^{c})\, \ell_{cv}\mone,
\qquad
(\Dr \tl_{cv'}-\Dr \tl_{cv}) =-\tl_{cv'}\,(\Dr L_{v'v}^{c'})\, \tl_{cv'}\mone.
\ee
By adding a vanishing  contribution $\la \Dr \ell_{cv'} \curlywedge   H_{cc'}  \Dr \tl_{cv}   H_{cc'} \mone\ra - \la\Dr \ell_{cv'} \curlywedge   H_{cc'}  \Dr \tl_{cv}   H_{cc'} \mone\ra$ to  \eqref{omdiscrete}, we obtain  
  \bes
-2\Omega_{cc'}
&=&  
\la (\Dr \ell_{cv'}-\Dr \ell_{cv}) \wedge   (\Dr H_{cc'} + H_{cc'}  \Dr \tl_{cv}   H_{cc'} \mone)\ra
 \nn\\ &+&
   \la\Dr   H_{c'c} \curlywedge (\Dr \tl_{cv'}-\Dr \tl_{cv})    \ra
  +\la \Dr \ell_{cv'} \wedge   H_{cc'}  (\Dr \tl_{cv'}-\Dr \tl_{cv})   H_{cc'} \mone\ra  \nn\\
&=& 
\la \,(\Dr L^c_{vv'})\, \curlywedge   \ell_{cv}\mone(\Dr H_{cc'}+H_{cc'}  \Dr \tl_{cv}   H_{cc'})\ell_{cv} \ra  \nn\\ &-&
\la \tl_{cv'}\mone (\Dr   H_{c'c} + H_{c'c}\Dr \ell_{cv'} H_{cc'})\tl_{cv'}
 \curlywedge \,(\Dr L_{v'v}^{c'})\,    \ra
\ees 

We can now use the variational continuity equations  \eqref{varcont} at $x=v'$ and \eqref{varcont2} at $x=v$, to get the simple expression
 \bea
2 \Omega_{cc'} &= &  -
\la \,\Dr L^c_{vv'}\, \curlywedge    \Dr H_{cc'}^{v}  \ra  +
\la \Dr   H_{c'c}^{v'} \curlywedge \,\Dr L_{v'v}^{c'}\,    \ra\cr
 &=& \la \Dr H_{cc'}^{v} \curlywedge   \Dr L^c_{vv'}  \ra  +
\la \Dl   H_{cc'}^{v'} \curlywedge \,\Dl L_{vv'}^{c'}\,    \ra,
 \eea 
 which is the desired result.

 \subsection{Ribbon network as the classical version of the quantum group spin network}
Let us recall that we  consider a cellular decomposition $\Gamma^*$ of  the 2d manifold $\Sigma$. We denote $\Gamma$ the dual 2-complex, made of nodes, links and faces. Let us see how the  model is now built in terms of the discretized variables. We focus, in this section, on the Euclidean case with $\Lambda<0$, since the Iwasawa decomposition is global in this case.

\medskip

\begin{figure}[h]
\begin{center}
\includegraphics[scale=.35]{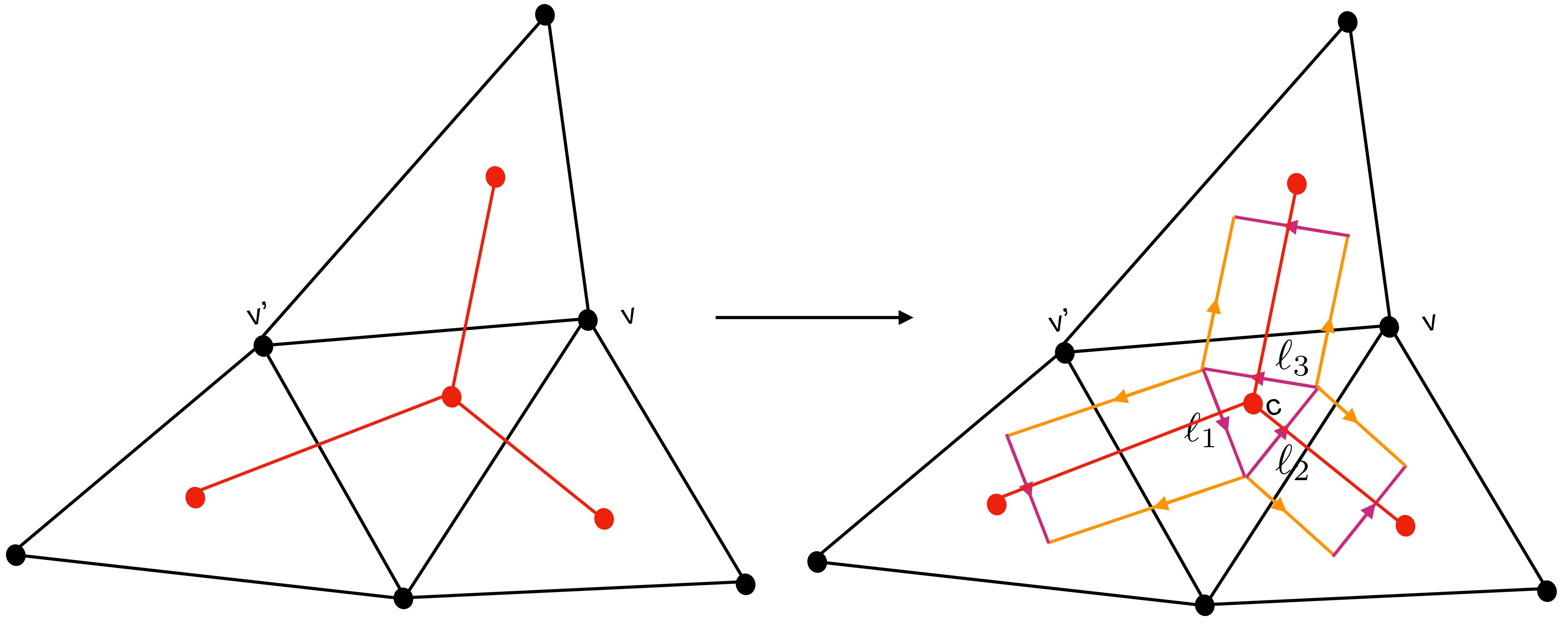}
\caption{The ribbon data encodes all the geometric data. In particular, when the ribbons meet at a node, the Gauss constraint $\ell_1\ell_2\ell_3=L^c_{v''v}L^c_{v'v''} L^c_{vv'}=1$ encodes the gauge invariance at the node and is the generalization of the flat case $X_1+X_2+X_3=0$.    }
\label{pic3}
\end{center}
\end{figure}

First the links are glued to each other at a node. For each link, we have a ribbon, hence we need to glue the ribbons together. By construction, the triangular holonomies in the $\AN$ sector going around a cell (eg a triangle) have a product which is the identity (as we assumed there is no torsion defect).  
\be
\cL^c= L^c_{vv'}L^c_{v'v''}L^c_{v''v}= \ell_{vc}\ell_{cv'} \ell_{v'c}\ell_{cv''} \ell_{v''c}\ell_{cv}=1.
\ee  
This indicates that the three ribbons ends form a closed $\AN$ holonomy and tells us how the ribbon are glued together,  see Fig. \ref{pic3}. This is the analogue of the Gauss constraint.

\medskip

Once the ribbons are glued together, we can also look at the faces generated by the``long side'' of the ribbon. Provided there is no curvature excitation, we expect to have a product of $\SU$ holonomies associated to the links $l_i$ around $v$ (or said otherwise the links which form the boundary of face $v^*$) being equal to the identity.  
\be
\cG^v= \prod_{l_i\in\partial v^*} \left(H^v_{l_i}\right)^{\pm1}=1,
\ee
where $\pm1$ depends on the orientation of the link $l_i$.

\medskip

These two sets of constraints provide the discretization of the (global) charges \eqref{new charges}.  As we have seen in Section \ref{sec:sym}, these two sets of holonomies generate the discrete analogue of the gauge transformations and the translations, as expected. Hence they should be seen as a discretization fo the charges $J'$ and $P'$ given in \eqref{new charges}, for constant transformation parameters on the boundary.  
Alternatively, one can check how the constraints $\cL^c$, $\cG^v$ can be viewed as a discretization of the generalized torsion and curvature constraints \eqref{gen tor}, \eqref{gen curv} (with no matter source).
{\fin
\begin{proposition}\label{prop:sym5}
The  holonomies $L^c_{vv'}$, $H^v_{cc'}$ are related to the (infinitesimal) continuum charges in the following way, with $h_{cx}$ a $\SU$ holonomy connecting $c$ to $x$ a point in the relevant path.
\bes
&&L^c_{vv'}= \cP \exp \left(\int_{[vv']} h_{cx} \act e(x) \right) , \\
&&H^v_{cc'}= \cP \exp \left(\int_{[vv']} \tom(x) - \left(h_{cx}^{-1}\, (h_{cx}\rhd e (x))\, h_{cx}\right)_{|_{\su}} \right)
\ees
The discrete constraints $\cL^c=L^c_{vv'} L^c_{v'v''} L^c_{v''v}=1$, $\cG^v=H^v_{cc'} .. H^v_{c^{(n)}c}=1$ encode that the generalized torsion and curvature are zero. 
\bes
 \cL ^c = 1 &\Leftrightarrow & \rd e + \tom\rhd e+\demi  [e \wedge e]_{\an}=0 \label{calc991}\\
\cG^v =1 &\Leftrightarrow&   
\rd \tom +  \tom \lhd e+\demi  [\tom \wedge \tom]_{\su}=0.
\ees
\end{proposition}
We leave the proof of the proposition in Appendix \ref{proofprop:sym5}. The expression of the discretized variables in terms of the continuum fields when $\Lambda\neq0$ is another aspect of the main result of this paper. }

 \medskip
 
 The (generalized) LQG phase space is  given in terms of the product of phase space $\DS_{\sigma s}^{l_i} $  associated to  the links $l_i=[c_ic_{i+1}]$, quotiented by the action of the (Gauss)  constraints $ \cL^{c^i}\equiv \prod_{j} L^{c^i}_{v_jv_{j+1}}$ acting at the nodes $c_i$. 
 \be
 \cP:= \times_i \DS_{\sigma s}^{l_i} /\!/ \cL^{c_i}
 \ee
The dynamics is given in terms of the contraints $\cG^{v}$ associated to the vertices $v_i$ of $\Gamma^*$, expressed in terms of the $H^{v_i}$.

%
%

\medskip

This model is exactly the model discussed in \cite{Bonzom:2014wva}.  The ribbon structure was proposed  to define the classical phase space structure of 3d gravity in the presence of a cosmological constant. Here this model is derived rigorously from the continuum. Note that \cite{Dupuis:2019pi} analyzed how such model can be related to the Fock-Rosly approach to the Chern-Simons formulation (in the case of the torus space).

\section{Recovering the quantum group structure}\label{qgp}
The quantum theory associated to the Heisenberg double phase space in the $\SU_{\sigma s}$ case is a  standard construction leading to the appearance of quantum group \cite{Chari:1994pz}. For the sake of being complete let us recall the  construction without going through all the technical details (see also \cite{Bonzom:2014bua} in the $\SU(2)$ case). Again, we focus on $\DS_{+-}=\SL(2,\C)= \SU(2)\bowtie \AN$, the Euclidean case with $\Lambda<0$. 

Constructing a quantum theory means that we use a representation of the relevant symmetries, which we saw in Sections \ref{sec:new action} and \ref{sec:sym} were associated to charges. In the case of 3d gravity, we have two types of symmetries, the rotation symmetries and the translations. While in the full theory  we need to implement both, the order in which we implement them at the quantum level matters. The different options are  first the rotations then the translations, or vice versa, or both at the same time. The first approach consists in the LQG picture,  the second one is "dual LQG" \cite{Delcamp:2018sef}, and the third one is the Chern-Simons  picture. 

In the following we will focus on the LQG approach, meaning that we will implement the rotational symmetry first, encoded by  the Gauss charges. 

\subsection{Poisson-Lie symmetry}

Before proceeding to quantization we need to tie one lose end.
The relationship between the  $r$-matrix $r$ entering the Poisson brackets  and the
 $r$-matrix $\mr$ entering the deformation of the action.
These are given by 
\be
r_- = -\tau_I \ot \bfJ^I\in \an \ot \su, \qquad 
\mr= \mr_{IJ} \bfJ^I\ot \bfJ^J \in \su \ot \su.
\ee
We have seen in \eqref{Heisc} that the Poisson brackets of the rotational holonomies is given by 
\bes
\poi{h_1,h_2}= -[\Rm,h_1h_2].
\ees
We expect however that the charge of symmetries acting on our phase space to belong the 
the Poisson-Lie group SU. This possesses the Poisson commutation relations 
\be 
\poi{h_1,h_2}= [\mr ,h_1h_2].
\ee
There seems to be a tension between this two results.
This tension is simply resolved by the  fact that these two expressions are the same.
\be\label{equivalence}
 [\mr ,h_1h_2] = -[\Rm,h_1h_2]
\ee
Strikingly this shows that  the $r$-matrix we have introduced at the very beginning as a boundary term \eqref{rmatbdy} enters as a structure constant deforming the symmmetry group action.  $\mr$  is the standard $r$-matrix encoding the deformation of the group $\SU(2)$ \cite{Semenov1992}, \cite{Chari:1994pz}. \textit{Our construction highlights that the notion of quantum group appears from the addition of the specific boundary term  in \eqref{rmatbdy}.}

One first establish it at the level of the Lie algebra:
Given $\alpha \in \su$ we want to prove that 
\be \label{inftsmal}
[\mr ,\alpha_1+\alpha_2] = -[\Rm,\alpha_1+\alpha_2].
\ee
This can be established by a direct computation as shown in \cite{Chari:1994pz}.
For the reader's convenience we present  it here explicitly.
Taking $\alpha=\bfJ^I$, and using \eqref{crosscom} and \eqref{Pcom}, we have 
\bea
-[\Rm,\bfJ^I\ot 1 +1\ot \bfJ^I]&=& [\tau_J,\bfJ^I]\ot \bfJ^J - \tau_J\ot [\bfJ^J,\bfJ^I] \cr
&=& -(C_{JK}{}^I \bfJ^K +\epsilon^{I}{}_{J}{}^K \tau_K)\ot \bfJ^J -
\tau_J\ot ( \epsilon^{JI}{}_{K} \bfJ^K),\cr
&=& C_{JK}{}^I \bfJ^J \otimes \bfJ^K \cr
&=& \sigma [(n \cdot \bfJ) \otimes \bfJ^I-\bfJ^I\ot (n \cdot \bfJ) ].
\eea
while on the other hand we have 
\bea
[\mr ,\bfJ^I\ot 1 +1\ot \bfJ^I] &=&\mr_{AB} \left( [\bfJ^A,\bfJ^I]\ot \bfJ^B + 
\bfJ^A\ot [\bfJ^B,\bfJ^I]\right) \cr
&=&\left( \mr_{AB} \epsilon^{AIC}  \right) ( \bfJ_C \ot \bfJ^B-\bfJ^B \ot \bfJ_C)\cr
&=& n^D \epsilon_{DAB}\epsilon^{AIC}( \bfJ_C \ot \bfJ^B-\bfJ^B \ot \bfJ_C)\cr
&=& \sigma n^D (\delta_{B}^I\delta_D^C-\delta_{B}^C\delta_D^I)( \bfJ_C \ot \bfJ^B-\bfJ^B \ot \bfJ_C)\cr
&=& \sigma  [ (n\cdot \bfJ) \ot \bfJ^I-\bfJ^I \ot (n\cdot \bfJ) ].
\eea
This establishes \eqref{inftsmal}.
The identity \eqref{equivalence} follows by exponentiation.

\medskip

\subsection{Quantization}

\paragraph{Specific representation choice. } It is useful to choose a specific representation to make some explicit calculations.
 An element $\ell$ in $\AN$ will be specified by a real number $\lambda$ and a complex number $z$.
\be
\ell \equiv  \begin{pmatrix} \lambda & 0 \\ z & \lambda^{-1} \end{pmatrix}, \quad \bar{\ell} \equiv \begin{pmatrix} \lambda\mone & -\overline{z} \\ 0 & \lambda \end{pmatrix}.
\ee
Note however that this representation is not faithful for $\AN$ so this is why we need to consider also $ \bar{\ell} \equiv \ell^{\dagger -1}$. (The map $G\to\bar{G}={G}^{\dagger -1}$ is a group morphism of $\AN$, which leaves the rotation subgroup invariant, as can be seen from the Iwasawa decomposition $\ell h\dr h^\dagger \ell^\dagger \dr \ell^{\dagger -1} h^{\dagger -1}=\bar{\ell} h$ \cite{Bonzom:2014bua}.) {\fin It is convenient to consider dimensionless Lie algebra generators, $(\sigma^I, \xi_I = i \sigma_I + ( \sigma \times \hat n)_I)$, where $\hat n=(0,0,1)$ is the (dimensionless) normalized vector, and   $\sigma_I$ are the (hermitian) Pauli matrices\footnote{Rescaled by a factor $\demi$.} with $[\sigma_I,\sigma_J]=i\,\epsilon_{IJ}{}^K\sigma_K$.  
 \be
\bfJ^I = -i \kappa \, \sigma^I, \quad \tau_I =  i \sqrt{|\Lambda|} \xi_I.
\ee
This means in particular that the $r$-matrix parametrizing the Poisson brackets will have an explicit parameter dependence (not hidden in the Lie algebra generators anymore as in section \ref{sec:sym}), given by $\mk= \ka \sqrt{|\Lambda|}$.  
This leads to an explicit expression for the $r$-matrix.  
\be
\Rm \,=\,- \tau^I\ot \bfJ_I\, =  -\, \mk \, \xi ^I\ot \sigma_I = i \frac{\mk}{4} \begin{pmatrix}- 1 &0 &0 &0\\ 0 &1 &0&0 \\ 0 &-4 &1 &0 \\ 0 &0 &0 &-1 \end{pmatrix}. 
\ee
 We recall that for a given link, we have the ribbon variables $\ell\in\AN$, with Poisson brackets
\bes\label{poirel}
&\poi{\ell_1,\ell_2}=\com{\Rm, \ell_1\ell_2}, \quad \poi{\bar\ell_1 ,\bar\ell_2 }=\com{\Rm, \bar\ell_1 \bar\ell_2 }, \quad  \poi{\ell_1 ,\bar\ell_2 }=\com{\Rm, \ell_1 \bar\ell_2 } .
\ees  

These are  equivalent to the following Poisson commutation relations 
\be
\poi{\lambda, z} = i\frac{\gamma}2 {z}\lambda ,\qquad \poi{\lambda, \bar{z} } = -i\frac{\gamma}2 {\bar{z}}\lambda,\qquad 
\poi{\bar{z},z} =- i\gamma( \lambda^2-\lambda^{-2}).
\ee
while other commutators vanish.

\paragraph{Quantization. }
Let us  quantize the matrix elements of $\ell$, so that they become operators  \cite{STERN_1995, Bonzom:2014bua}. 
We first introduce the parameter  
\be
q= e^{\hbar\gamma/2}, \textrm{ with } \hbar\gamma= \hbar \kappa \sqrt{|\Lambda|} = 8 \pi \frac{ l_P}{l_c},
\ee 
where  $l_P=\hbar G$ is the Planck length and $l_C=|\Lambda|^{-\demi}$ is the cosmological scale. 

We define then the deformed quantum monodromy matrix 
\be
\ell 
\dr \hat \ell =\begin{pmatrix} K & 0 \\ (q - q^{-1}) J_+ & K^{-1} \end{pmatrix}, \quad \bar\ell 
\dr \hat {\bar\ell} =\begin{pmatrix} K\mone & -(q - q^{-1}) J_-  \\ 0 & K \end{pmatrix} ,
\ee
where the correspondence is 
\be 
 K= \hat\lambda ,\quad
 (q-q^{-1}) J_+ = \hat{z},\qquad
 -(q-q^{-1}) J_- = \hat{\bar{z}}.
 \ee
%
The classical $r$-matrix becomes the quantum $R$-matrix 
\be\label{quR}
\Rm \dr R_-=  q^{-\frac12} \begin{pmatrix} q & 0 & 0 & 0\\ 0 & 1 & 0 & 0\\ 0 &  (q - q^{-1}) & 1 & 0\\ 0 & 0 & 0 & q \end{pmatrix}= \one +i \hbar\,\Rm + \mathcal{O}(\hbar^2). 
\ee
 Finally the Poisson brackets which appears though the limit $[\hat{\ell}_1,\hat{\ell}_2] \to -i \hbar\poi{{\ell}_1,{\ell}_2}$, 
are quantized through  
\bes \label{RLL=LLR}
R_-\,\hat{\ell}_1\,\hat{\ell}_2 = \hat{\ell}_2\,\hat{\ell}_1\,R_- &\dr&  \poi{\ell_1,\ell_2}= 
\com{\Rm, \ell_1\ell_2}, \nn\\ 
R_-\,\hat{\bar\ell}_1\, \hat{\bar\ell}_2 = \hat {\bar\ell}_2\,\hat{\bar\ell}_1\,R_- &\dr&
  \poi{{\bar\ell}_1 , {\bar\ell}_2 }=\com{\Rm,  {\bar\ell}_1  {\bar\ell}_2 },  \nn\\ 
R_-\,\hat{\ell}_1\,\hat{\bar\ell}_2 = \hat{\bar\ell}_2\,\hat{\ell}_1\,R_- &\dr&
 \poi{\ell_1 ,{\bar\ell}_2 }=\com{\Rm , \ell_1 {\bar\ell}_2 }.
\ees
In components, the commutation relations on the right hand side of  \eqref{RLL=LLR} read}
\be
K\,J_+\,K^{-1} = q\,J_+,\qquad K\,J_-\,K^{-1} = q^{-1}\,J_-,\qquad [J_+,J_-] = \frac{K^2 - K^{-2}}{q - q^{-1}}.
\ee
These are the commutation relations of $\mathcal{U}_q(\SU(2))$. This is encoding the well-known fact that the quantum algebra of functions on $\AN$ is isomorphic to the algebra $\mathcal{U}_q(\SU(2))$. 

%

\smallskip

The last element we need is the  Hilbert space. Since we intend  first to implement  the rotational symmetries,  we consider the natural Hilbert space  associated to the $\hat \ell$ which actually span $\mathcal{U}_q(\SU(2))$. Hence we consider the Hilbert space given in terms of the irreducible representations of  $\mathcal{U}_q(\SU(2))$. Strictly speaking we should consider such Hilbert space for a half link, and glue two of such representations to build a full link as  recalled in \cite{Delcamp:2018sef}. We will skip these subtleties here.

\smallskip

Now that we have the quantum theory for a given link, we need to extend the structure to the full graph $\Gamma$. For simplicity we have taken $\Gamma^*$ to be a triangulation so that the nodes of $\Gamma$ are trivalent. For each node, we have three $\AN$ holonomies, belonging to different phase spaces, which product is $1$. This is the Gauss law. The product is given by the matrix product. 

The quantum version of the Gauss law is direct. Since we have to consider three phase spaces, we have to deal with three Hilbert space copies, with each quantum $\AN$ holonomy acting a given Hilbert space. The $\AN$ holonomies are multiplied using the matrix product, hence the natural quantization  of the holonomy product is
\be
(\ell\ell')_{ik} = \sum_j(\ell)_{ij}(\ell')_{jk} \dr (\widehat{\ell\ell'})_{ik} \equiv \sum_j(\widehat\ell)_{ij}\ot (\widehat\ell)_{jk}= \cop \widehat \ell_{ik} . 
\ee
This is nothing else than the natural coproduct for the algebra of functions on $\AN$. 
We read in terms of the components,
\be
\begin{aligned}
&\cop \hell
= \begin{pmatrix} K\otimes K & 0\\ (q-q^{-1}) (J_{+} \otimes K + K^{-1}\otimes J_+) & K^{-1}\otimes K^{-1} \end{pmatrix},\\
\text{and}\qquad &\cop \hat{ \bar \ell} 
= \begin{pmatrix} K\otimes K &  -(q-q^{-1}) (J_{-} \otimes K + K^{-1}\otimes J_-)\\ 0 & K^{-1}\otimes K^{-1} \end{pmatrix}.
\end{aligned}
\ee
We recognize the coproduct of $\mathcal{U}_q(\SU(2))$. The Gauss constraint demanding that the product of the three $\AN$ holonomies is 1 is then quantized as 
\be
 1_{ik}= (\ell\ell'\ell'')_{ik} = \sum_{jl}\ell_{ij}\ell'_{jl}\ell''_{lk}   \dr  \sum_{jl}\widehat \ell_{ij} \ot\widehat\ell_{jl}\ot \widehat\ell_{lk}  = (1\ot \cop)\circ \cop \widehat\ell_{ik}= \widehat1_{ik}.
\ee
The elements in the Hilbert space solutions of such constraints are the $\mathcal{U}_q(\SU(2))$ intertwiners, generated by the deformed Clebsh-Gordan coefficients.
%
 We recover in this way the $\mathcal{U}_q(\SU(2))$ spin networks. Solving then the last set of constraints for the $\SU(2)$ holonomies gives rise to the Turaev-Viro amplitude\footnote{The TV model is usually defined for $\cU_q(\su(2))$ with  $q$ root of unity  to have a finite model. The other signature and cosmological constant sign cases usually lead to a divergent model, just like the Ponzano-Regge model. These divergences can be understood as signaling the presence of a non-compact symmetry and can be gauged away \cite{Freidel:2002dw}. } \cite{Turaev:1992hq}.

\section*{Outlook}
In this work we investigated why, at the quantum level,  a deformed gauge symmetry, parametrized by the cosmological constant $\Lambda$, appears whereas the original action for 3d gravity is a plain undeformed gauge theory.  

\smallskip

The first key insight was  to realize that we had to perform a  change of variables at the continuum level, in order to have a Gauss constraint/rotational charge algebra depending upon the cosmological constant.  The change of variables is a simple canonical transformation parametrized by a  vector $n$ which equivalently can be seen as induced by  a boundary term. Such vector $n$ is taken as a scalar (ie an invariant) for the gauge symmetries and therefore leads to a modification of the realization of the symmetries. Since $n$ is constrained to depend on $\Lambda$, we do get symmetries that depend on $\Lambda$ at the action level.   

This is yet another example that the  choice of variables matters  in the quest of defining a proper quantum gravity theory. There is an obvious parallel in our work and the 4d LQG approach where one performs a canonical transformation parameterized by a scalar, the Immirzi parameter or equivalently   adds a (topological) term not modifying the equations of motion, the Holst term to define the Ashtekar-Barbero variables. This canonical transformation renders the theory more amenable to discretization, just like our term does for 3d gravity. The main difference however is that $n$ is parameterized by $\Lambda$ so it is not really adding an extra parameter in the theory unlike the Immirzi parameter.

\smallskip 

The second key insight is the discretization procedure. It is in fact a  subtle procedure: we have decomposed the system into subsystems and managed to project all the degrees of freedom on the boundary of the subsystems by imposing an appropriate truncation of the degrees of freedom. Such truncation is obtained by going on-shell. In the 3d gravity context, this amounts to consider region of homogeneous curvature and no torsion. This is essentially the same as   dealing with the notion of "geometric structures"\cite{Carlip:1998uc}  or equivalently homogeneously curved polygons.  A boundary shared by two polygons can be viewed from the perspective of each polygon, and an isometry relating the two,  the so-called  continuity equations. This allowed to express the discrete variables solely in terms of "corner" terms (the classical version of the Kitaev triangle operators \cite{Kitaev1997} \cite{triangle}). {\fin From this perspective, the quantum group symmetry appears in a sense as the "corner term contributions". 
Note also  that our work shares some similarities with the seminal works  \cite{Gawedzki:1990jc}\cite{Falceto_1993}, where the quantum group symmetry is identified at the classical level for the Wess-Zumino model. }

\smallskip

The phase space associated to each link of the graph $\Gamma$ (dual to the triangulation $\Gamma^*$) now depends on the cosmological constant $\Lambda$. Importantly, we have derived this phase space (the Heisenberg double)  starting from the continuum symplectic form. It was already known that   such Heisenberg double equipped with the appropriate constraints,  provides a discretization of 3d gravity with a non-vanishing cosmological constant \cite{Bonzom:2014wva} and also leads upon quantization to deformed spin networks and the TV amplitude \cite{Bonzom:2014bua}.  We have therefore found the missing link connecting the discretized model and the continuum model. This paper provides therefore   a long thought-for and rigorous derivation of the quantum group structure -- as a kinematical symmetry-- in the 3d loop quantum gravity case. {\fin Interestingly it can also provide the link between the Fock-Rosly approach and the gravity continuum variables, since it was explicitly shown in \cite{Dupuis:2019pi}  how such approach was related to the ribbon model \cite{Bonzom:2014wva}.}

\smallskip

This works opens many new avenues of investigation. Let us review some of them. 

\paragraph{More general vector $n$.}
There is some room to go beyond the quantum group case, by removing some conditions on the vector $n$,
  \be
\delta n =0, \quad n^2=-\Lambda^2, \quad \rd n=0, \quad n^I= (0,0,n^3).
\ee
We can consider for example a  vector such $\rd n\neq0$, which would generate some new central extension \eqref{newcom1} that would be interesting to explore.  

In our construction, the vector $n$ is a scalar for the symmetries, with its norm fixed by the cosmological constant. Hence in a sense, the only relevant information we keep about $n$ is its norm. It would be interesting to see how its direction could also be relevant. 
For example,  two vectors $n$, related by a  rotation lead to isomorphic quantum group structures. At the classical level a rotation of the vectors corresponds to a canonical transformation. It would be interesting to see whether this is the case at the quantum group level.
That is it possible to relate explicitly two rotated quantum group structure by a unitary transformation?

\paragraph{Unexplored cases.} For the sake of simplicity, we focused on the simplest cases. Indeed as we argued earlier, the   Euclidean case with positive cosmological constant has to be treated separately due to appearance of  reality conditions since we have to deal with a complex $n$. 

In the Lorentzian cases, we focused on the component connected to the identity to use the Iwasawa decomposition, $\ell h = \th\tl$, but one should deal with the general case, where there exist $d_i, d_j\in\ds_{-,s}$, such that {\fin $\ell d_i h = \th d_j\mone\tl$. }The Heisenberg double can be generalized accordingly \cite{Alekseev_1994}. This amounts however to decorate the ribbon by some curvature parametrized by  $d_i$.

We have studied only one polarization choice in the discretization in section \ref{sec:main}. Namely we looked at the case where $\AN$ holonomies are associated to the edges whereas the $\SU$ holonomies are associated to the links. Due to the symmetric treatment between the two groups, we can actually swap the location of the holonomies. 
{\fin In fact the continuity equation   \eqref{intcont} also allows  to identify the dual  variables.
\bes
&&G_{c v}G_{v c'}=  G_{c v'}G_{v' c'}
\Leftrightarrow \quad \tilde L^v_{cc'} \tilde H^{c'}_{vv'}=  \tilde H^{c}_{vv'}\tilde L^{v'}_{cc'}.
\ees
Hence we have $\tilde L^x_{cc'}\in\AN$, with $x$ being $v$ or $v'$, associated to the links and $ \tilde H^{y}_{vv'}\in\SU$, with $y$ being $c$ or $c'$,  associated to the edges.}
This   provides a deformation of the dual loop formalism \cite{Dupuis:2017otn, Shoshany:2019ymo}, which should be  the classical analogue of  \cite{Dittrich:2016typ} (for the case $q$ real though).  
We leave the study of this other polarization for later studies.

It is clear that our construction can be generalized to any factorizable group. Namely, considering  a BF theory associated with a simple Lie group $G$, we expect the boundary deformation to  be given in terms of the standard $r$-matrix and the main  results and proofs to generalize seamlessly. We leave this for future work.  

\paragraph{Adding matter.}
While we did not introduce matter, in the shape of curvature or torsion excitations, the formalism can certainly be extended to this case. We expect that  the edge mode (or corner terms) perspective provides naturally the notion of particles in the curved case, just like it did in the flat case \cite{Freidel:2018pbr}. 


We expect then to recover a version of the Kitaev model, defined for (deformation of) Lie groups. It would be then interesting to explore how much gravity questions we could ask in the Kitaev model context. This would develop some new interplay between models of (topological) quantum information theory and quantum gravity.


\paragraph{Recent progresses.}
Following the publication of this article, a number of significant developments have taken place.

For example the edge mode/covariant program for gravity, which we used,  was more systemically addressed in 4d \cite{Freidel2020_edgesI,Freidel2020_edgesII}. The symplectic transformation induced by the boundary term described in Section \ref{sec:canon} has been extended to any dimension \cite{Girelli:2021pol}. In particular, it was used to show how to define a generalization of teleparallel gravity in the presence of a cosmological constant. A relation with the Henneaux--Teitelboim model for unimodular gravity was also identified in this context.  

A covariant phase space/edge mode  approach was similarly used to explore very specific 4d BF theories \cite{Girelli:2021zmt}. While the notion of Heisenberg double does not exist yet for 2-groups in full generality (see however \cite{Girelli:2021khhs} for some attempt in this direction), the notion of 2-Drinfeld classical double was identified for BF theories \cite{Chen:2022opt}. These are the first steps to have a complete analogue treatment for 4d BF theory as the one we did here.       

Considerations about the shrinkability of some (inner) boundary/entangling surface  also show that an effective $q$-deformation of the symmetries \cite{Mertens2023_factorization} must appear.  Interestingly, this deformation is actually different  than the one derived here.  A better understanding of why this is the case will certainly further highlight how the notion of (gauge) symmetries is very much intertwined with the notion of boundary.

 \subsection*{Acknowledgements}
\noindent  This research was supported in part by Perimeter Institute for Theoretical
Physics. Research at Perimeter Institute is supported by the Government
of Canada through the Department of Innovation, Science and Economic
Development Canada  and by the Province of Ontario through the Ministry
of Research, Innovation and Science. A.\,O.\ is supported by the NSERC Discovery grants held by M.\,D.\ and F.\,G.

\appendix
\section{Playing with cross and dot products}\label{sec:cross}

\subsection{Poisson algebra of charges}\label{proof1}
We explicitly  calculate the Poisson bracket between the different charges generating the deformed symmetries. We work with $\ka=1$.
\bea
\{{P'}_\alpha,P'_\beta\} &=& 
\{ P_\alpha + J_{\alpha\times n} , P_\beta + J_{\beta\times n}\}\cr
&=& \sigma\Lambda J_{\alpha\times \beta } +
P_{ (\alpha\times n)\times \beta - (\beta\times n) \times \alpha}
+ J_{(\alpha\times n)\times (\beta\times n)}  \quad {\scriptscriptstyle \alpha\times (\beta\times n)- \beta\times (\alpha \times n)= (\alpha \times \beta)\times n} 
\cr
&=& (\sigma\Lambda +\sigma n^2)  J_{\alpha\times \beta}  +
P_{ (\alpha\times \beta) \times n }  + J_{((\alpha\times \beta) \times n )\times n}
\quad {\scriptscriptstyle (\alpha\times n)\times (\beta\times n)
=((\alpha \times \beta)\times n)\times n 
+ \sigma n^2 (\alpha \times \beta)}
\cr
&=&  (\sigma\Lambda +\sigma n^2)  J'_{\alpha\times \beta } +  P'_{ (\alpha\times \beta) \times n } 
\cr
\{J'_\alpha,P'_\beta\} &=& 
\{ J_\alpha  , P_\beta + J_{\beta\times n}\}\cr
&=& 
P_{ \alpha\times \beta   }  + J_{\alpha  \times ( \beta \times n ) }\cr
&=&  {P}_{ \alpha\times \beta   }  + J_{\beta  \times ( \alpha \times n ) }
+ J_{(\alpha\times \beta)\times n} \quad {\scriptscriptstyle \alpha\times (\beta\times n)- \beta\times (\alpha \times n)
= (\alpha \times \beta)\times n }\cr
&=&  P'_{ \alpha\times \beta   }  + J'_{\beta \times (\alpha \times n)} 
\eea

\subsection{$\ds_{\sigma s}$ gauge theory.}\label{big-guy}
Let us consider the $\ds$ connection $\cA=\tom+e$, then the curvature of $\cA$ is 
\bes
F[\cA]&=& \rd \cA + \demi [\cA\wedge \cA]= \rd \tom + \rd e +\demi [(\tom+e)\wedge (\tom+e)]\nn\\
&=& \rd \tom + \rd e + \demi [\tom\wedge \tom]+\demi [e \wedge e] +   [\tom\wedge e]\nn\\
&=& \rd \tom + \rd e + \demi [\tom\wedge \tom]+\demi [e \wedge e] +   \tom\rhd  e + \tom \lhd e\nn\\
&=&( \rd \tom+\demi [\tom\wedge \tom]+  \tom \lhd e ) + (\rd e +\demi [e \wedge e] +   \tom\rhd  e )= \mF + \mT,
\ees
which is the sum of the generalized curvature $\mF$ in the $\su$ direction and the generalized torsion $\mT$ in   the $\an$ sector.

To determine the derivative in the different sectors, we consider the $\ds_{\sigma s}$ element $\psi=\alpha+\phi$, $\alpha\in\su$  and $\phi\in\an$,
\bes
\rd_\cA \psi &=& \rd \alpha+\rd \phi + [(\omega+ e), (\alpha+\phi)] \\
  &=& (\rd \alpha + [ \omega, \alpha] + \omega \lhd \phi + e \rhd \alpha) + (\rd \phi+ [e,\phi]+   \omega \rhd \phi + e \lhd \alpha)
\ees
Setting either of $\alpha$ or $\phi$ to be zero, we get the derivative in the respective directions.
\bes
D \alpha &=&\rd \alpha + [ \omega, \alpha]_\su +  e \rhd \alpha \\
&=& \rd \alpha +  \omega \times \alpha  +   e\times (n\times \alpha) \\
\tilde D \phi &=& \rd \phi+ [e,\phi]_\an+   \omega \rhd \phi \\
&=&\rd \phi+  (e\times \phi)\times n+   \omega \times \phi 
\ees
These derivative satisfy the metric compatibility condition
\be
\rd  (\alpha \cdot \phi) =  D\alpha \cdot \phi + \alpha \cdot \tilde{D}\phi,
\ee
which can be shown directly  using the definition of the vector triple product $(\alpha\times \beta)\times n =\sigma[ (\alpha\cdot n)\beta -\alpha (n\cdot \beta) ]. 
$

We can evaluate the generalized Bianchi identities that the different components satisfy from the Bianchi identity of the big curvature $\cF$.
\bes
\rd_\cA\cF=0\dr\left\{ \begin{array}{l} D \mF + \omega \lhd \mT =0\\
\tilde D \mT + \mF\rhd e=0  \end{array}\right. 
\ees

\section{Some proofs}
\subsection{Proof of Proposition \ref{prop:sym1}} \label{proofprop:sym1}
We want to prove that 
\bes
\delta^L_\alpha   \ip  \Omega_{c c'}
&=& \la \alpha  \, , \, \Dr \ell \ra 
\ees
 It is actually only necessary to use \eqref{whererot}  in the symplectic form to identify the charge generating the infinitesimal rotations. 
The following identities will be useful to do the proof. 
\bes
 \th\mone \ell h =  \tell &\Leftrightarrow& \Dl \tell = h \mone(\ell\mone \Dl \th  \ell + \Dl  \ell +  \Dr h  )h\nn
\\
 \la h\mone \delta^L_\alpha h\,,\,    \Dl \tell\ra&=&  \la \delta^L_\alpha h h\mone\,,\,   (\ell \mone \Dl \th\mone  \ell + \Dl  \ell +  \Dr h   )\ra \nn\\
 &=&  \la  \delta^L_\alpha hh\mone \,,\,   \ell {}\mone ( -\Dr h   + \Dr  \ell   )\ell \ra \nn
\ees
With this in mind the calculation is direct. 
\bes
\delta^L_\alpha   \ip  \Omega_{c c'}&=& \demi \left(\la \delta^L_\alpha \th \th\mone \, ,\,  \Dr \ell \ra + \la   \delta^L_\alpha h h \mone\,,\,    h \Dl \tell h \mone\ra -  \la \Dr \th\, ,\,  \delta^L_\alpha \ell \ell\mone \ra  \right)\nn\\
&=& \demi \left(\la \delta^L_\alpha \th \th \mone \, ,\,  \Dr \ell \ra + \la  \delta^L_\alpha hh\mone \,,\,   \ell\mone ( -\Dr \th  + \Dr  \ell   ) \ell\ra 
-  \la \Dr \th\, ,\,  \delta^L_\alpha \ell \ell\mone \ra  \right)\nn\\
&=&\demi \left(\la \delta^L_\alpha \th\th\mone +  \ell \delta^L_\alpha hh \mone   \ell \mone \, , \, \Dr \ell\ra 
-   \la   \delta^L_\alpha \ell\ell\mone  + \ell  \delta^L_\alpha hh\mone    \ell \mone  \, ,\, \Dr \th   \ra  \right)\nn\\
&=&\demi \left(\la \alpha + \delta^L_\alpha G G\mone -   \delta^L_\alpha \ell\ell\mone \, , \, \Dr \ell \ra -   \la    \delta^L_\alpha G G\mone  \, ,\, \Dr \th  \ra  \right)\nn \\
&=&\la \alpha  \, , \, \Dr \ell\ra -   \demi\la    \alpha \, ,\, \Dr \th   \ra  \nn\\
&=& \la \alpha  \, , \, \Dr \ell \ra 
\ees

\subsection{Proof of Proposition \ref{prop:sym2}} \label{proofprop:sym2}
We want to prove that the Poisson brackets 
\bea
\poi{\ell_{1},\ell_{2}}= \com{\Rm, \ell_{1}\ell_{2}}, &\, &
\poi{\ell_{1},h_{2}}=\ell_{1} \Rm h_{2}, \,\, 
 \\ 
\poi{\ell_{1}, \tell_{2}}= 0, &\, &
 \,\, 
\poi{\ell_{1}, \th_{2}}= \Rm \, \ell_{1} \th_{2}. \nn
 \eea
 with $\Rm= - \tau_I \ot  \bfJ^I$
are the right brackets to generate the infinitesimal transformations, through the formula 
\be
\delta^L_\alpha  \cdot  =  -\la \alpha_1+\phi_1 \,,\,\{\ell_{1}\,,\, \cdot    \} \ell_{1}\mone \ra_{1}, \ee
where $\alpha \in \su$ and $ \phi \in \an$.
The fact that $\phi$ is projected out is necessary to interpret 
$\{\ell_{1}\,,\, \cdot     \} \ell_{1}\mone$ as a vector field in AN.

The proof goes as follows
\bes
 -\la \alpha_1+\phi_1\,,\,\{\ell_{1}\,,\, \ell_{2}     \} \ell_1\mone \ra_{1} &=&  -\la \alpha_1+\phi_1\,,\,\com{\Rm, \ell_{1}\ell_{2}} \ell_1\mone \ra_{1} \cr
 &=& \la (\alpha+\phi),\tau_I\ra  \bfJ^I \ell - 
 \la (\alpha+\phi), \ell \tau_I \ell\mone \ra \ell  \bfJ^I \nn\\
&=& \alpha \ell -\la \ell\mone \alpha \ell \,,\,  \tau_I \ra \ell  \bfJ^I =  \alpha \ell - (  \alpha \lhd \ell)_I\ell  \bfJ^I   \nn\\
&=&  \alpha \ell - \ell (  \alpha \lhd \ell)=\delta^L_\alpha \ell ,
\ees
where we used that $\alpha \lhd \ell =  (\ell\mone \alpha \ell)_{|_\su} =\la \ell\mone \alpha \ell \,,\,  \tau_I \ra   \bfJ^I$ and that $\la \ell\mone \phi \ell \,,\,  \tau_I \ra=0$.

Similarly, taking     
\bes
  -\la \alpha_1 +\phi_1 \,,\,\{\ell_{1}\,,\, h_{2}     \} \ell_{1}\mone \ra_{1} &=&   -\la \alpha_1+\phi_1\,,\,\ell_{1}\, \Rm \,h_{2} \ell_{1}\mone \ra_{1} \cr
  &=& \la \alpha +\phi\,,\, \ell \tau_I \ell\mone\ra   \bfJ^I h \nn\\
&=&  ( \ell\mone \alpha \ell)_{|_{\su}}   h= (  \alpha \lhd \ell)h 
=\delta^L_\alpha h
\ees
Finally 
\bes
 -\la \alpha_1+\phi_1\,,\,\{\ell_{1}\,,\, \th_{2}     \} \ell_{1}\mone \ra_{1}  &=&  
 - \la \alpha\,,\,\Rm \,  \th_{2} \ra_{1} \cr
 &=& \la \alpha+\phi\,,\,  \tau_I \ra   \bfJ^I \th  =  \alpha    h= \delta^L_\alpha \th,
\ees
while 
\bes
\poi{\ell_{1}, \tell_{2}}=0 &\dr &\delta^L_\alpha \tell =0.
\ees
which completes the proof.

\subsection{Proof of Proposition \ref{prop:sym4}} \label{proofprop:sym4}
We want to prove that the Poisson brackets 
\bes
 \poi{\th_{1},\th_{2}}=\com{\Rp,\th_{1}\th_{2}}, \quad
\poi{\th_{1}, \tell_{2}}= \th_1\Rp \tell_2, \quad  
\poi{\th_{1}\,,\,h_{2}}=0, \quad 
\poi{\th_{1}\,,\,\ell_{2}} = \Rp \th_{1} \ell_{2}. 
\ees
are the right brackets to generate the infinitesimal transformations, through the formula 
\be\label{babelon21}
\delta^L_\phi \cdot = \la \alpha_1+\phi_1\,,\,\{\th_{1}\,,\, \cdot     \} \th_{1}\mone \ra_{1}, 
\ee
where $\alpha \in \su$ and $ \phi \in \an$.

The proof goes as follows. First
\bes
\delta^L_\phi \th &=& \la \alpha_1+\phi_1\,,\,\{\th_{1}\,,\, \th_{2}     \} \th_{1}\mone \ra_{1} = \la \alpha_1+\phi_1\,,\,\com{\Rp,\th_{1}\th_{2}} \th_{1}\mone \ra_{1}  \nn\\
&=&\la \alpha_1+\phi_1\,,\, \bfJ_I\ra \tau^I \th -  \la \alpha_1+\phi_1\,, \th\bfJ_I\th\mone \ra  \th \tau^I  = \phi \th - \th  (\th\mone \phi \th)_{|_\an} \nn\\
&=&\phi \th - \th  ( \phi  \lhd \th ).
\ees
Then the other proofs are direct. 
\bes
\poi{\th_{1}, \tell_{2}}= \th_1\Rp \tell_2 &\dr& \delta^L_\phi \tell   (\th \mone \phi \th)_{|_\an} \tell  = ( \phi \lhd \th ) \th\\
\poi{\th_{1}\,,\,h_{2}}=0 &\dr & \delta^L_\phi h=0 \\
\poi{\th_{1}\,,\,\ell_{2}} = \Rp \th_{1} \ell_{2} &\dr & \delta^L_\phi \ell = \phi \ell. 
\ees

\subsection{Proof of Proposition \ref{prop:sym5}} \label{proofprop:sym5}

First we want to find the relation between the discrete charges and the continuum ones. 
Let us consider the $\AN$ holonomy $L^c_{vv'}= \ell_{vc}\ell_{cv'}$. It is enough to focus on the single holonomy $\ell_{cx}$ for $x\in[vv']$, as in Fig. \ref{loop04}. We can express $\ell_{cx}$ in terms of the $\an$ connection $\ove (y) \equiv \ell_{cy}\mone \rd \ell_{cy}$. 
\be
\ell_{cx}= \cP \exp \left(\int_{cx} \ove \right).
\ee
In a similar way, we can define a $\SU$ holonomy $h$ and connection $\ovo (y) \equiv h_{cy}\mone \rd h_{cy}$.  
\be
h_{cx}= \cP \exp \left(\int_{cx}\ovo \right).
\ee
 The connections $\ovo, \ove$ are actually related to the spin connection $\tom$ and frame field $e$. Recall that we took  in \eqref{con1}, \eqref{con2}, omitting the subscripts $cy$,
\bes 
\tom^I\bfJ_I  &\equiv&  h^{-1} \text{d}h \: + \: \left(h^{-1} (\ell^{-1} \text{d}\ell) h\right)_{|_{\su}}\label{con11} = \ovo\: + \: \left(h^{-1}\, \ove\, h\right)_{|_{\su}} \\
  e_I\tau_I  &\equiv&   \left(h^{-1} \, (\ell^{-1} \text{d}\ell)\, h\right)_{|_{\an}}  =\left(h^{-1} \, \ove\, h\right)_{|_{\an}}  \equiv h\mone \act \ove\, . \label{con22}
\ees 
The action we defined $h \act \ove= \left(h \, \ove\, h\mone\right)_{|_{\an}} $ is indeed an action since 
\bes
g \left(h \, \ove\, h\mone\right)g\mone &=& \left((g h) \, \ove\, (gh)\mone \right)_{|_{\an}}+\left((g h) \, \ove\, (gh)\mone \right)_{|_{\su}} = (gh)\rhd \ove +\left((g h) \, \ove\, (gh)\mone \right)_{|_{\su}} \nn\\
&=& g \left(h \, \ove\, h\mone\right)_{|_{\an}} g\mone + g \left(h \, \ove\, h\mone\right)_{|_{\su}} g\mone = (g \left(h \, \ove\, h\mone\right)_{|_{\an}} g\mone)_{|_{\an}} + (g \left(h \, \ove\, h\mone\right)_{|_{\an}} g\mone)_{|_{\su}} \nn\\ &&+ (g \left(h \, \ove\, h\mone\right)_{|_{\su}} g\mone) _{|_{\su}} \nn\\
\Leftrightarrow (gh)\rhd \ove&=& (g \left(h \, \ove\, h\mone\right)_{|_{\an}} g\mone)_{|_{\an}} = 
g\rhd (h\rhd \ove).\nn
\ees
Now we deduce that 
\bes
\ove(x)= h_{cx}\rhd e (x), \quad \ovo(x)= \tom(x) - \left(h_{cx}^{-1}\, \ove(x)\, h_{cx}\right)_{|_{\su}}.
\ees
This allows to have explicitly that 
\bes
L^c_{vv'}&=& \ell_{vc}\ell_{cv'}= \cP \exp \left(\int_{cc'} h_{cx}\rhd e (x)  \right)
,\\
H^v_{cc'}&=& h_{cv}h_{vc'}=
\cP \exp \left(\int_{vv'} \tom(x) - \left(h_{cx}^{-1}\, (h_{cx}\rhd e (x))\, h_{cx}\right)_{|_{\su}} \right)
\ees

\smallskip
 
We want to find the infinitesimal constraints behind the discrete Gauss and flatness constraints. Let us first focus on the Gauss constraint. 
\bes
&&\cL^c =\prod_i \ell^c_i=1 \Leftrightarrow \rd \ove + \demi [\ove\wedge\ove]_{\an}=0. \label{r1}
\ees
To determine what is \eqref{r1} in terms of the frame field $e$ and the connection $\tom$, we first identify that $\ove = \left(h \, e\, h^{-1} \right)_{|_{\an}} $ from \eqref{con22}. 
We will use the identities coming from the match pair properties  
\bes
\left(h [h\mone \rd h, e]h\mone \right)_{|_{\an}}&=& \left(h [h\mone \rd h, e]_{|_{\an}} h\mone \right)_{|_{\an}}= h\rhd [h\mone \rd h, e]_{|_{\an}}= h\rhd (( h\mone \rd h)\rhd e) , \nn\\
h\mone [\ove\wedge\ove]h&=& h\mone\rhd [\ove\wedge\ove]_\an + (h\mone [\ove\wedge\ove]h)_{|_\su}\nn\\
&=&  [(h\mone\ove h)\wedge(h\mone\ove h)]=  [(h\mone\rhd \ove )\wedge(h\mone\rhd \ove)]_\an + 2 (h\mone\ove h)_{|_\su}\act (h\mone\ove h)_{|_\an} \nn\\
&&+ 2 (h\mone\ove h )_{|_\an}\act (h\mone\ove h)_{|_\su}  + [(h\mone\ove h ){|_\su}\wedge(h\mone\ove h){|_\su}]_\su \nn\\
\Leftrightarrow  h\mone\rhd [\ove\wedge\ove]_\an &=& [(h\mone\rhd \ove )\wedge(h\mone\rhd \ove)]_\an + 2 (h\mone\ove h)_{|_\su}\act (h\mone\ove h)_{|_\an} \nn \\
&=& [e\wedge e]_\an + 2 (h\mone\ove h )_{|_\su}\act e  \nn
\ees
Plugging the expression of $\ove$ in \eqref{r1}, we get 
\bes
0=\rd \ove + \demi [\ove\wedge\ove]_{\an}&=& h\rhd \rd e + \left(h [h\mone \rd h, e]h\mone \right)_{|_{\an}}+ \demi [\ove\wedge\ove]_{\an}\nn\\
&=&  h \rhd \rd e + h \rhd (h\mone \rd h\rhd  e)+ \demi [\ove\wedge\ove]_{\an}\nn\\
&=&  h \rhd \rd e + h \rhd ((\tom\: - \: \left(h^{-1}\, \ove\, h\right)_{|_{\su}})\rhd  e)+\demi h\rhd [e \wedge e]_{\an}+h\rhd ((h\mone\ove h )_{|_\su}\act e )  \nn\\
&=&  h \rhd (\rd e + \tom\rhd e+\demi  [e \wedge e]_{\an}) \label{calc99}
\ees
This is the deformed continuous Gauss constraint \eqref{defGauss}. 

\smallskip

Nest we want to prove that 
\be
F[\ovo]= \rd \ovo  + \demi [\ovo \wedge\ovo ]_{\su}=0 \Leftrightarrow \rd \tom +\demi [\tom \wedge \tom ]_\su +  \omega\lhd e =0.
\ee
As before a number of identities are necessary to prove to get the equivlance. First, denoting $[,]$ for the Lie algebra $\ds$ bracket, we have 
\bes
\demi h\mone \, [\ove \wedge \ove ]_\an \, h &=& \demi h\mone\,  [\ove \wedge \ove ] \, h =  \demi  [ h\mone \ove h\, \wedge h\mone\ove\, h ] \nn\\
&=& \demi [ (h\mone \ove h)_{|_\an}\,  \wedge \, (h\mone \ove h)_{|_\an}]+ [ (h\mone \ove h)_{|_\su}\,  \wedge \, (h\mone \ove h)_{|_\an}]\nn\\ 
&&+ \demi [ (h\mone \ove h)_{|_\su}\,  \wedge \, (h\mone \ove h)_{|_\su}] \nn
\ees
This means that we have 
\bes 
\demi (h\mone\, [\ove \wedge \ove ]_\an\, h)_{|_\su} =  [ (h\mone \ove h)_{|_\su}\,  \wedge \, (h\mone \ove h)_{|_\an}]_{|_\su} + \demi [ (h\mone \ove h)_{|_\su}\,  \wedge \, (h\mone \ove h)_{|_\su}] _\su .\label{plouf1}
\ees
We also check that  
\bes
\rd (h\mone \ove h)_{|_\su}&=& (h\mone \,\rd \ove\, h)_{|_\su} - [(h\mone \ove h)\wedge \ovo]_{|_\su}\nn\\
&=&-\demi  (h\mone [\ove\wedge\ove ]_\an h)_{|_\su} - [(h\mone \ove h)_{|_\su}\wedge \ovo]_{|_\su}- [(h\mone \ove h)_{|_\an}\wedge \ovo]_{|_\su}\label{plouf2}\\
&=&- [ (h\mone \ove h)_{|_\su}\,  \wedge \, (h\mone \ove h)_{|_\an}]_{|_\su} - \demi [ (h\mone \ove h)_{|_\su}\,  \wedge \, (h\mone \ove h)_{|_\su}] _\su \nn\\
&-&  [(h\mone \ove h)_{|_\su}\wedge (\tom - \left(h^{-1}\, \ove\, h\right)_{|_{\su}} ) ]_{|_\su}- [(h\mone \ove h)_{|_\an}\wedge (\tom - \left(h^{-1}\, \ove\, h\right)_{|_{\su}} )]_{|_\su}\nn \\
&=& \demi [ (h\mone \ove h)_{|_\su}\,  \wedge \, (h\mone \ove h)_{|_\su}] _\su -  [(h\mone \ove h)_{|_\su}\wedge \tom   ]_{|_\su}- [(h\mone \ove h)_{|_\an}\wedge \tom]_{|_\su}\nn \\
&=& \demi [ (h\mone \ove h)_{|_\su}\,  \wedge \, (h\mone \ove h)_{|_\su}] _\su -  [(h\mone \ove h)_{|_\su}\wedge \tom   ]_{|_\su}- \omega\lhd e   \label{plouf3},
\ees
where in \eqref{plouf2} we used \eqref{plouf1}, and in \eqref{plouf3}, we used the definition of the frame field, as well as the definition of the action of $\an$ on $\su$. This means that 
\bes
\rd_\omega  (h\mone \ove h)_{|_\su}&=& \rd (h\mone \ove h)_{|_\su} + [\omega\wedge (h\mone \ove h)_{|_\su}  ]_\su \nn\\
&=& \demi [ (h\mone \ove h)_{|_\su}\,  \wedge \, (h\mone \ove h)_{|_\su}] _\su -  [(h\mone \ove h)_{|_\su}\wedge \tom   ]_{|_\su}- \omega\lhd e   + [\omega\wedge (h\mone \ove h)_{|_\su}  ]_\su \nn\\
&=& \demi [ (h\mone \ove h)_{|_\su}\,  \wedge \, (h\mone \ove h)_{|_\su}] _\su - \omega\lhd e   \label{plouf4}.    
\ees
With this in mind, the relation between $F[\ovo]$ and the generalized curvature is direct.  Recalling that $\ovo= \tom - (h\mone \ove h)_{|_\su}$, 
\bes
F[\ovo] &=& F[\tom] - \rd_\tom  (h\mone \ove h)_{|_\su} + \demi [(h\mone \ove h)_{|_\su}\wedge (h\mone \ove h)_{|_\su}]\nn\\
&=&F[\tom] + \omega\lhd e ,
\ees
where we just replaced the value of $\rd_\omega  (h\mone \ove h)_{|_\su}$ determined in \eqref{plouf4}.




\end{document}